\documentclass[aps,pre,preprint,showpacs]{revtex4}
\usepackage{graphicx}
\usepackage{amssymb}
\usepackage{epsfig}
\usepackage{subfigure}
\usepackage{amsmath}

\begin{document}

%=========================================================

\title{\bf Phase and Amplitude Dynamics in Large Systems of Coupled
           Oscillators: Growth Heterogeneity, Nonlinear Frequency Shifts and Cluster States}

\author{Wai Shing Lee, Edward Ott and Thomas M. Antonsen Jr.
	%\\
	%\texttt{wslee@umd.edu}
       }

\affiliation{Institute for Research in Electronics and Applied Physics, 
                  University of Maryland, College Park, Maryland 20742, USA}

%==========================================================

\begin{abstract}
This paper addresses the behavior of large systems of heterogeneous, 
globally coupled oscillators each of which is described by the generic Landau-Stuart 
equation, which incorporates both phase and amplitude dynamics of individual oscillators.
One goal of our paper is to investigate the effect of a spread in the 
amplitude growth parameter of the oscillators and of the effect of a homogeneous nonlinear frequency
shift. Both of these effects are of potential relevance to recently reported
experiments. Our second goal is to gain further understanding of the macroscopic system dynamics
at large coupling strength, and its dependence on the nonlinear frequency shift parameter.
It is proven that at large coupling strength, if the nonlinear frequency shift parameter
is below a certain value, then there is a unique attractor for which the oscillators all
clump at a single amplitude and uniformly rotating phase
(we call this a single-cluster ``locked state''). Using a combination of analytical and
numerical methods, we show that at higher values of the nonlinear frequency shift parameter,
the single-cluster locked state attractor continues to exist, but other types of coexisting 
attractors emerge. These include two-cluster locked states, periodic orbits, chaotic orbits,
and quasiperiodic orbits.

\end{abstract}

\maketitle

{\bf Systems of coupled oscillators occur in a very wide variety of applications. Often interaction
between the dynamical evolution of the oscillator phases and amplitudes is 
an important issue.
The simplest model of such dynamics is that of a globally coupled system
of Landau-Stuart equations \cite{PRK}. While this system is very basic, due to the large space
of possibilities for its parameters and their probability distribution functions, 
a complete understanding of the system is lacking. Here, motivated
by recent experiments \cite{Taylor}, we consider parameter dependences 
not previously investigated. We also investigate the reason for the
common occurrence of ``locked states" (constant amplitude and sinusoidal
oscillation) observed in previous studies when the coupling is large, 
why non-locked-state attractors
occur at sufficiently large values of the nonlinear frequency shift parameter,
and what types of attractors can occur at large coupling and large nonlinear frequency
shift.}

%============================================================ intro
\section{Introduction} \label{sec:intro}
The interaction of many coupled dynamical units is a common
theme across a broad range of scientific disciplines. Within 
this general theme, the issue of determining conditions for
the emergence of macroscopic cooperative behavior and of determining
the nature of this behavior is of central
importance \cite{PRK,Strogatz1}. Examples include coupled lasers
\cite{Kozyreff}-\cite{Zamora}, Josephson junction circuits
\cite{Marvel,Nichols}, interacting yeast cells \cite{Dano,Monte},
pacemaker cells in the heart \cite{Michaels}, pedestrian 
induced oscillation of footbridges \cite{MillB}-\cite{Abdulrehem},
chemically reacting systems \cite{Kiss,Taylor}, circadian
rhythms \cite{Yamaguchi}, and many others.

A very useful simplified
framework for beginning to understanding phenomena observed in
these situations is the phase oscillator description.
In the phase oscillator description the dynamical units are assumed
to be oscillatory with fixed amplitude. Thus, the state of
each individual unit can be specified solely by a phase angle
$\theta$, and the evolution of oscillator $i$ is taken to be 
determined by its present state $\theta_i$ and by the states
$\theta_j$ of the other oscillators ($j \neq i$). The simplest
model of this type was originally put forward by Kuramoto in 1975 and has
proven to be an extremely useful paradigm for understanding 
this general type of system \cite{Kuramoto}-\cite{Ott}. In addition,
the Kuramoto model has also served as a basis for formulating related
phase oscillator models appropriate to a wide variety of situations
(e.g., see Ref. \cite{OA}). The basic question addressed by the Kuramoto
model is the competition between the synchronizing effect of coupling
and the desynchronizing effect of different natural frequencies
of the individual oscillators. The principal result \cite{Kuramoto}-\cite{OA}
coming from the solution of the Kuramoto model is that,
in the limit of a large number of oscillators, this competition is 
resolved by a transition, whereby there is a critical coupling
strength below which the oscillations of individual oscillators occur
with random phase and there is no macroscopic population-wide oscillation,
while above which oscillators begin to develop phase coherence, and 
globally-averaged population-wide oscillation emerges.

A drawback of the phase oscillator description is that, by its definition,
it excludes the effect of amplitude dynamics and the possible
coupling of amplitude dynamics with phase dynamics. Another
useful oscillator model incorporating both amplitude and phase dynamics is based
on the normal form of an isolated system near a Hopf bifurcation,

% --------------------------------------------------------- LanStu
\begin{equation}
\frac{dz}{dt} = (\alpha + i \omega) z - (\beta + i\gamma) |z|^2 z,
\label{eq:LanStu}
\end{equation}

\noindent
also referred to as a Landau-Stuart oscillator \cite{PRK}. In 
(\ref{eq:LanStu}) $z$ is complex with $|z|$ and the angle of $z$
representing the amplitude and phase of the oscillator. The real
parameter
$\alpha$ is the linear amplitude growth rate of
the oscillations, with $\alpha > 0$ for growth (and $\alpha < 0$ for 
damping). The Hopf bifurcation occurs as $\alpha$ passes through 
zero. The other
real parameters $\omega, \beta, \gamma$ respectively
characterize the small-amplitude natural frequency of the oscillator,
and the finite amplitude nonlinear shifts of the small amplitude
growth rate and frequency. The bifurcation is supercritical if
$\beta > 0$ (the nonlinear term saturates growth)
and subcritical (hysteretic) if $\beta < 0$ (the nonlinear term enhances growth). 
Here we will
only deal with the supercritical case [in the subcritical case,
if $\alpha > 0$,
orbits typically go far from $z=0$, thus
invalidating the expansion resulting in (\ref{eq:LanStu})]. For
$\alpha < 0$, Eq. (\ref{eq:LanStu}) has as its stable solution
$z=0$. For $\alpha > 0$, $z=0$ is unstable, and (\ref{eq:LanStu})
results in an attracting limit cycle attractor,

% --------------------------------------------------------- LanStu_limcy
\begin{equation}
z = \sqrt{\frac{\alpha}{\beta}} \exp \left[i \left(\omega-
\frac{\gamma \alpha}{\beta} \right) t + \theta_0  \right],
\label{eq:LanStu_limcy}
\end{equation}

\noindent
which traces a circular orbit about the origin 
of the complex $z$-plane.
In general, the normal form
oscillator parameters ($\alpha,\omega,\beta,\gamma$) derived from the
physical system under study will depend on some set, 
${\bf p}=( p^{(1)},p^{(2)}, \cdots,p^{(n)} )^T$, of physical parameters
for that system. That is, $[\omega,\alpha,\beta,\gamma]=
[\omega({\bf p}),\alpha({\bf p}),\beta({\bf p}),
\gamma({\bf p})]$. 

We are interested in the situation, also
studied in Refs. \cite{Shiino}-\cite{Pazo}, where many oscillators of the
form of Eq. (\ref{eq:LanStu}) are coupled together and where each
such oscillator (indexed by a subscript $i=1,2, \cdots, N \gg 1$) may have
a different parameter set. That is, if oscillator $i$ has parameter
set ${\bf p}_i$, then  

% ----------------------------------------------------------- para
\begin{equation}
[\omega_i,\alpha_i,\beta_i,\gamma_i]=
[\omega({\bf p}_i),\alpha({\bf p}_i),\beta({\bf p}_i),
\gamma({\bf p}_i)]. 
\label{eq:para}
\end{equation}

\noindent
If the value of 
${\bf p}$ is regarded as assigned
randomly from oscillator to oscillator according to some
probability distribution function (pdf), then that will induce a corresponding
pdf $\hat{G}$ of the parameters 
$[\omega,\alpha,\beta,\gamma]$, such that 

% ------------------------------------------------------------ hatG
\begin{equation}
\hat{G}(\omega,\alpha,\beta,\gamma) d\omega d\alpha d\beta d\gamma
\label{eq:hatG}
\end{equation}

\noindent
represents the fraction of oscillators with parameters
$\omega,\alpha,\beta,\gamma$ in the range 
$\omega \in [\omega,\omega+d\omega]$,
$\alpha \in [\alpha,\alpha+d\alpha]$,
$\beta  \in [\beta,\beta+d\beta]$, 
$\gamma \in [\gamma,\gamma+d\gamma]$,
and applicable in the limit 
$N \rightarrow \infty$, where $N$ is the number of oscillators.

Considering this general problem, one would like to know how the system behavior
depends on the distribution function $\hat{G}(\omega,\alpha,\beta,\gamma)$.
However, as $\hat{G}$ is a distribution in the four variables 
$\omega, \alpha, \beta, \gamma,$ this is clearly too big a problem
to address in full generality. Here we will pursue a more modest 
program. In particular, the questions we address are partly motivated by the experimental 
work in Ref. \cite{Taylor}:
(i) what is the effect of spread in $\alpha$ allowing the simultaneous
presence of dead ($\alpha < 0$) and active ($\alpha > 0$) oscillators in the
uncoupled state, and (ii) what is the effect of a nonlinear frequency shift $\gamma$
(for simplicity, we treat the oscillators as all having the same $\gamma$),
and iii) how we can understand certain types of simple nonlinear behavior often observed
in these systems when the coupling strength between oscillators is large?

% ========================================================== formulation

\section{Formulation, background and outline} \label{sec:formulation}
We assume  
$\beta$ and $\gamma$ are the same for all oscillators, $\beta_i=\bar{\beta}$
and $\gamma_i = \bar{\gamma}$. Furthermore, we scale $\bar{\beta}$ to one
by a proper normalization of $z_i$ ($z_i \rightarrow z_i / \sqrt{\bar{\beta}}$).
Thus

% -------------------------------------------------------------- hatG2
\begin{equation}
\hat{G}(\omega,\alpha,\beta,\gamma) = G(\omega,\alpha)  
                          \delta(\beta-1) \delta(\gamma - \bar{\gamma}).
\label{eq:hatG2}
\end{equation}

\noindent
If $\omega$ and $\alpha$ are uncorrelated in their variation from
oscillator to oscillator, then $G$ is of the form 

% -------------------------------------------------------------- Gfact
\begin{equation}
G(\omega,\alpha) = g(\omega) h(\alpha).
\label{eq:Gfact}
\end{equation}

\noindent
In what follows we assume that Eq. (\ref{eq:Gfact}) holds \cite{note1},
and that $g(\omega)$ is symmetric and monotonically decreasing with
respect to its maximum value, which we can take to be located at 
$\omega=0$ (if the maximum of $g(\omega)$ occurred at some non-zero 
value, $\omega = \bar{\omega}$, then the location of the maximum can be
shifted to zero by the change of variables $z=z'e^{i \bar{\omega}t}$,
$\omega'=\omega-\bar{\omega}$).

For (\ref{eq:LanStu}) with $\beta_i=1$, $\gamma_i=\bar{\gamma}$ and 
(\ref{eq:Gfact}) specifying our ensemble of uncoupled oscillators, we now
proceed to globally couple these ensemble members through a mean 
field, $\langle z \rangle$,

% ------------------------------------------------------------- LanStuCou
\begin{subequations}
\begin{align}
\frac{dz_i}{dt} &= (\alpha_i + i \omega_i) z_i - 
           (1 + i\bar{\gamma}) |z_i|^2 z_i + \Gamma \langle z \rangle,
           \label{eq:LanStuCou1} \\
\langle z \rangle &= \frac{1}{N} \sum_{j=1}^N z_j, \label{eq:LanStuCou2}
\end{align}
\label{eq:LanStuCou}
\end{subequations}

\noindent
where the parameter $\Gamma$ measures the strength of the coupling and is assumed 
real and positive, $\Gamma \geq 0$ (some previous studies have considered complex
coupling constants, e.g., Refs. \cite{Hakim}-\cite{Nakagawa2}). 
We will sometimes refer to $\langle z \rangle$
as the ``order parameter'' because whether or not there is global
collective behavior for $N \rightarrow \infty$ corresponds to 
whether $|\langle z \rangle| >0$ or $\langle z \rangle=0$. See
Refs. \cite{Shiino}-\cite{Pazo} for previous related work on 
large coupled systems of Landau-Stuart equations. In many of these
previous works \cite{Shiino}-\cite{Matthews2}, the coupling term is written
as $K(\langle z \rangle - z_i)$ in place of $\Gamma \langle z \rangle $.
This choice is simply related to ours by the transformation 
$\alpha_i = \hat{\alpha}_i - K$, $\Gamma = K$. We prefer the parametrization
in Eq. (\ref{eq:LanStuCou}) because one of our principal motivations will be 
experiments \cite{Taylor} where it can be plausibly argued that quantities
analogous to $\Gamma$ and the average value of $\alpha_i$
(denoted $\bar{\alpha}$) can be varied essentially independently.
More generally, in real large coupled oscillator systems
familiar to us, coupling between the oscillators typically
results from physical processes distinct from those determining
the properties of the individual oscillators
(as in Ref. \cite{Taylor}), and the parametrization in 
Eq. (\ref{eq:LanStuCou}) is therefore the appropriate one. Use of the
form (\ref{eq:LanStuCou}) (rather than (\ref{eq:LanStu_oldw}) below)
will be important for our considerations of the large coupling
limit in Sec. \ref{sec:LS_strong_couple}.
In addition, in Refs. \cite{Shiino}-\cite{Matthews2} it was 
considered that $\hat{\alpha}_i$ was the same positive constant
for all $i$, $\hat{\alpha}_i=\hat{\alpha}$, and 
furthermore that $\bar{\gamma}=0$. 
Parameter and time
normalizations were then chosen to transform $\hat{\alpha}$ to $1$,
yielding, in place of (\ref{eq:LanStuCou}) 

% ---------------------------------------------------------- LanStu_oldw
\begin{equation}
\frac{dz_i}{dt} = (i \omega_i + 1 - |z_i|^2) z_i + 
           K (\langle z \rangle - z_i),
\label{eq:LanStu_oldw}
\end{equation}

\noindent
where the coupling parametrization form $K(\langle z \rangle - z_i)$ was used.
We, however, will be interested in the effect of a spread in 
$\alpha_i$ with the possibility of the simultaneous occurrence 
of positive and negative $\alpha_i$ for different $i$, and also
in the effects of nonlinear frequency shift $\bar{\gamma} \neq 0$.

One motivation for this study is the recent paper,
Ref. \cite{Taylor}, which describes experiments in which
many ($\sim 10^{4} / cm^3$) specially designed small porous
particles are continuously and rapidly mixed in a 
catalyst-free Belousov-Zhabotinsky reaction mixture. The catalyst
for the reaction is immobilized on the small porous particles, each of which
can potentially serve as an effective chemical oscillator. 
Oscillations in the chemical states of the particles are 
visualized as the color of the particles oscillates between red and blue. The particle
density serves as a parameter analogous to our coupling constant 
$\Gamma$, while regulation of
the stirring rate effectively provides a control 
analogous to control of the mean oscillator growth rate,

% ---------------------------------------------------------- meanalp
\begin{equation}
\bar{\alpha} = \int \alpha h(\alpha) d\alpha.
\label{eq:meanalp}
\end{equation}

\begin{figure}[h]
  {\includegraphics[width=6.1cm]{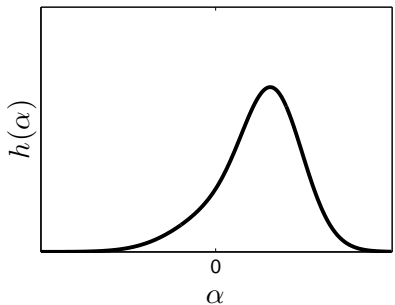}} 
  \caption{Schematics of $h(\alpha)$.}
  \label{fig:sch_h}
\end{figure}

\noindent
Because the process by which the particles are prepared is not
perfect, it is expected that there will be substantial
spread in their parameters, and in particular in $\omega$
and $\alpha$. These spreads are of particular interest because:
(i) spread of oscillator frequencies is the essential feature leading
to the transition from incoherently oscillating units to macroscopic
oscillation in the Kuramoto model, and (ii) the parameter 
$\alpha$ determines whether individual particles, when uncoupled,
oscillate ($\alpha > 0$) or do not oscillate ($\alpha < 0$). 
In the case $\alpha < 0$ the attractor for Eq. (\ref{eq:LanStu}) 
is the fixed point $z=0$, often referred to as ``oscillator death''.
With reference to point (ii), because of the spread
in $\alpha$, in some range of stirring rates, we can expect a situation
like that shown schematically in Fig. \ref{fig:sch_h}, 
which depicts an uncoupled oscillator growth rate pdf $h(\alpha)$ yielding substantial
fractions of the particles in the oscillating and dead states.
As $\bar{\alpha}$ increases from very negative values, 
$(-\bar{\alpha}) \gg \delta \alpha)$ (analogous to low stirring rates in the
experiment), to very positive values, $\bar{\alpha} \gg \delta \alpha$
(analogous to high stirring rates), there is a continuous transition from
predominantly dead to predominantly oscillatory dynamics of the uncoupled
oscillators. Another notable feature of these experiments is that 
the collective coherent frequency of oscillation exhibits a marked dependence
on the oscillation amplitude through its dependence on the 
density of the porous particles at fixed stirring rate (e.g., the third panels in Fig. 2(a)
and 2(b) of Ref. \cite{Taylor}). This is a strong indication that the nonlinear
frequency shift $\bar{\gamma}$ plays a significant role. It is notable
that Ref. \cite{Taylor} developed a set of chemical rate 
equations that, when solved numerically, yield good agreement with
the experiments. While this is a singular achievement, we are interested
in obtaining additional understanding of the processes involved and 
in determining if it is generic. To the extent that qualitative behavior of our 
Landau-Stuart model mimics behavior observed in the particular 
experiment in Ref. \cite{Taylor}, the typicality of the observed
phenomena is strongly implied. Furthermore, if the above
agreement holds, then any analytical results obtained for the 
Landau-Stuart model may lead to further
understanding of these experimental phenomena. Thus it is our desire to employ the
generic coupled Landau-Stuart model, Eqs. (\ref{eq:LanStuCou}),
to explore and understand the nature of the interplay 
between frequency spread, growth rate spread and nonlinear frequency shift.
In this connection, it is worth noting that our work may be applicable 
to other experiments. Indeed, as described in Ref. \cite{Taylor},
the chemical experiment was, at least partly, intended to mimic
observed oscillator quorum-sensing in yeast populations
\cite{Dano,Monte}. In addition, the basic stability analysis technique
used here (Sect. \ref{sec:linsta}) is similar to that originally introduced in 
Refs. \cite{Matthews} and \cite{Matthews2} can also be applied 
to other amplitude / phase oscillator systems, such as the laser system
considered in Ref. \cite{Zamora}.

% fffffffffffffffffffffffffffffffffff

We now give a brief review of the most important papers \cite{Shiino}-\cite{Pazo}
related to our work. References \cite{Shiino}-\cite{Matthews2}
considered Eq. (\ref{eq:LanStu_oldw}) (all oscillators have identical
$\alpha_i$ and $\gamma_i=0$) and examined the behavior
as a function of the
coupling constant $K$ and the spread $\sigma$ in the oscillator frequencies.
Shiino and Frankowicz \cite{Shiino} by a combination of numerical experiments
and analysis obtain an approximate $K-\sigma$ plane phase diagram. References
\cite{Ermentrout,Mirollo} examine the transition between
``amplitude death'' \cite{note2,Suarez} (i.e., $z_i=0$ for all oscillators)
and collective oscillation, explicitly obtaining analytical
results for the boundary in $K-\sigma$ space separating death and collective
oscillation.

Matthews et al. \cite{Matthews,Matthews2}, in addition to presenting an extensive
numerical exploration, also develop an analytical technique for handling 
the transition to globally coherent oscillation from phase-incoherent 
individual oscillation with $|z_i|>0$ (as in the Kuramoto transition
\cite{Kuramoto1}-\cite{OA}); thus this work was the first to include analysis of
the effect of amplitude dynamics on this type of transition. 
In addition, another
important result of Refs. \cite{Matthews, Matthews2} was the numerical discovery that
near the boundary in parameter space where the transition
to collective behavior occurs, this collective behavior can
be rather complex, including period doubling cascades,
chaos, quasiperiodicity and hysteresis. Further, sufficiently far above the boundary
it was found that steady oscillatory behavior prevails  (as in the 
Kuramoto model). 

Reference \cite{Daido} introduces a situation that
the authors
call ``aging'' in which there are two populations, each described by an equation
of the form of (\ref{eq:LanStuCou}) (with $\alpha_i=\hat{\alpha}-K$ and 
$\Gamma=K$); the ``old'' population has $\hat{\alpha}_i=-\hat{\alpha}_o < 0$
(corresponding to amplitude death at $K=0$), and the ``young'' population
has $\hat{\alpha}_i = \hat{\alpha}_y > 0$; $\omega_i$ was taken to be the same
constant $\Omega$ for all old and young oscillators (see also \cite{Pazo} 
which allows distinct old and young natural frequencies, $\omega_i=\Omega_o$ and
$\Omega_y$); and behavior was investigated as a function of the ratio of the 
populations of old relative to young. 
In the set up of Refs. \cite{Daido,Pazo},
due to the homogeneity of frequencies, the transition 
problem reduces to the analysis of two coupled Landau-Stuart equations. 

Nonlinear behavior of large systems of {\it identical} Landau-Stuart oscillators
was considered by Refs. \cite{Hakim}-\cite{Daido2}, which highlight the possible
occurrence of ``clustered states,'' such that oscillators in the same cluster
all behaves identically, but $z_i(t) \neq z_j(t)$ if $i$ and $j$ are in different clusters.

The rest of this paper is organized as follows. 
Section \ref{sec:linsta} derives the characteristic equation governing linear stability of
perturbations from the $\langle z \rangle = 0$ state. 
Section \ref{sec:Lorent_freq_dist} evaluates the characteristic equation for the case of a 
Lorentzian frequency distribution, $g(\omega)=[\pi (1+\omega^2)]^{-1}$.
Section \ref{sec:condlinsta} evaluates the effect of spread in $\alpha$ on linear stability
in the case of Lorentzian $g(\omega)$ and no nonlinear frequency shift ($\bar{\gamma}=0$).
Section \ref{sec:nonlinear_effect} evaluates the effect of nonlinear frequency shift ($\bar{\gamma} \neq 0$)
on stability in the case of Lorentzian $g(\omega)$ and no spread in $\alpha$. 
Section \ref{sec:freqdist_effect} considers a 
flat-top frequency distribution, $g(\omega)=U(1-|\omega|)/2$
(where $U(\bullet)$ denotes the unit step function) and investigates whether the 
qualitative behavior found in Secs. \ref{sec:condlinsta}
and \ref{sec:nonlinear_effect} is affected by this change in the form of $g(\omega)$.
Section \ref{sec:above_instab} discusses behavior above the instability threshold 
for cases when there is no spread in the nonlinear parameters $\beta$ and $\gamma$
of (\ref{eq:LanStu}) (as in Eq. (\ref{eq:LanStuCou})). 
Section \ref{sec:LS_strong_couple} studies stability of the corresponding 
nonlinear solutions in the limit of
large coupling, $\Gamma / \Gamma_c \rightarrow \infty$, where $\Gamma_c$ denotes the
critical value of $\Gamma$ at which the $\langle z \rangle = 0$ state becomes unstable.
A primary issue addressed in Secs. \ref{sec:above_instab} and \ref{sec:LS_strong_couple}  
is the explanation of why, for sufficiently small $\bar{\gamma}$, 
macroscopic solutions become purely 
oscillatory with constant amplitude as $\Gamma / \Gamma_c$ is increased (referred
to as the ``locked state"), and shows that 
multiple-clustered states with complex dynamics
can occur at large $\bar{\gamma}$.
Conclusions and further discussions are given in Sec. \ref{sec:conclusions}.

% ========================================== linsta  -- Derivation of the Dispersion Relation}
\section{Linear stability of the $\langle z \rangle = 0$ state} \label{sec:linsta}

We consider Eqs. (\ref{eq:LanStuCou}) in the limit $N \rightarrow \infty$.
In this case, there is a solution corresponding to zero value of the
order parameter $\langle z \rangle$. For $\langle z \rangle = 0$,
Eq. (\ref{eq:LanStuCou1}) has the solutions

% -------------------------------------------------------- incoh
\begin{subequations}
\begin{align}
z_i &= 0 \hspace{2mm} \mbox{for}  \hspace{2mm} \alpha_i < 0, \label{eq:incoh_1} \\
z_i &= \sqrt{\alpha_i} \exp\{ i [(\omega_i-\bar{\gamma}\alpha_i)t + \theta_{0i}] \}
 \hspace{2mm} \mbox{for}  \hspace{2mm} \alpha_i > 0. \label{eq:incoh_2}
\end{align}
\label{eq:incoh}
\end{subequations}

\noindent
We express the order parameter $\langle z \rangle$ as 
% ------------------------------------------------------- ordp
\begin{subequations}
\begin{align}
\langle z \rangle &= \langle z \rangle_- + \langle z \rangle_+, \label{eq:ordp_1}\\
\langle z \rangle_- &= \frac{1}{N} \sum_{i, \alpha_i < 0} z_i, \label{eq:ordp_2}\\
\langle z \rangle_+ &= \frac{1}{N} \sum_{i, \alpha_i > 0} z_i, \label{eq:ordp_3}
\end{align}
\label{eq:ordp}
\end{subequations}

\noindent
That is, $\langle z \rangle_-$ and  $\langle z \rangle_+$ denote the contribution
to the order parameter from oscillators with $\alpha_i < 0$ and 
$\alpha_i > 0$, respectively. Note that (\ref{eq:incoh_1}) implies
$\langle z \rangle_- = 0$, while (\ref{eq:incoh_2}) implies 
$\langle z \rangle_+ = 0$ if $N \rightarrow \infty$ and the angles
$\theta_{0i}$ are uniformly distributed in $[0,2\pi]$. Thus, by 
(\ref{eq:ordp_1}), we see that $\langle z \rangle = 0$ is indeed
a self-consistent solution of the system (\ref{eq:LanStuCou}) for
$N \rightarrow \infty$. We now ask whether this solution is stable 
to small perturbations. If it is not, then the state $\langle z \rangle = 0$
will not persist, and global collective behavior will result. We denote
the perturbation of the order parameter by

% -------------------------------------------------------------- ordp_pert
\begin{subequations}
\begin{align}
\langle \delta z \rangle &= \langle \delta z \rangle_- 
         + \langle \delta z \rangle_+, \label{eq:ordp_pert1}\\
\langle \delta z \rangle_{\pm} &= \frac{1}{N} \sum_{i,\alpha_i \gtrless 0} \delta z_i,
                               \label{eq:ordp_pert2}
\end{align}
\label{eq:ordp_pert}
\end{subequations}

\noindent
where $\delta z_i$ is a perturbation from the unperturbed orbit dynamics
given by Eqs. (\ref{eq:incoh}).

{\it Calculation of $\langle \delta z \rangle_-$}. Considering oscillator $i$ 
for which $\alpha_i < 0$, and perturbing Eq. (\ref{eq:LanStuCou1}) about
$z_i=0$, we obtain the linearized equation,

% ------------------------------------------------------------ LanStuCouN
\begin{equation}
\frac{d \delta z_i}{dt} = (\alpha_i + i \omega_i) \delta z_i + 
\Gamma \langle \delta z \rangle.
\label{eq:LanStuCouN}
\end{equation}

\noindent
Assuming exponential time dependence of the orbit perturbations, 
$\delta z_i \sim \exp(st)$, Eq. (\ref{eq:LanStuCouN}) yields

% ------------------------------------------------------------- incohN_pert
\begin{equation}
\delta z_i = \frac{\Gamma \langle \delta z \rangle}{(s + |\alpha_i| - i \omega)}
\hspace{2mm} \mbox {for} \hspace{2mm} \alpha_i < 0.
\label{eq:incohN_pert}
\end{equation}

\noindent
Thus
% ------------------------------------------------------------ incohN_pert_avg
\begin{equation}
\langle \delta z \rangle_- = \Gamma \langle \delta z \rangle \int_{-\infty}^{\infty}
\int_{-\infty}^{0} \frac{g(\omega)h(\alpha)}{(s + |\alpha| - i \omega)} 
d\alpha d\omega,
\label{eq:incohN_pert_avg}
\end{equation}

\noindent
where for $N \rightarrow \infty$ we have replaced the sum over $i$ in 
(\ref{eq:ordp_2}) by integration over $\omega$ and $\alpha$ weighted
by the pdf's $g(\omega)$ and $h(\alpha)$ [Eq. (\ref{eq:Gfact})].
Note that the $\alpha$ integration in (\ref{eq:incohN_pert_avg})
runs from $\alpha=-\infty$ to $\alpha=0$ and thus includes
only those oscillators for which $\alpha<0$.

{\it Formulation for calculating $\langle \delta z \rangle_+$}. We begin
by re-expressing Eq. (\ref{eq:LanStuCou1}) in polar form, 
$z_i=\rho_i \exp(i \theta_i)$ where $\rho_i(t)$ and $\theta_i(t)$
are real,

% --------------------------------------------------------------- LanStuCouP
\begin{subequations}
\begin{align}
\frac{d \rho_i}{dt} &= \alpha_i \rho_i - \rho_i^3 + \Gamma Re\{ e^{-i\theta} 
                       \langle z \rangle\}, \label{eq:LanStuCouP_rho} \\
\frac{d \theta_i}{dt}&= \omega_i - \bar{\gamma}\rho_i^2 + \frac{\Gamma}{\rho_i}
             Im \{ e^{-i \theta}  \langle z \rangle \}, \label{eq:LanStuCouP_the} \\
\langle z \rangle &= \langle \rho e^{i \theta} \rangle. \label{eq:LanStuCouP_3}
\end{align}
\label{eq:LanStuCouP}
\end{subequations}

\noindent
We now introduce a pdf for the state variables $(\rho,\theta)$ and parameters
$(\omega,\alpha)$ which we denote by

% ----------------------------------
\begin{equation*}
f(\rho,\theta,\omega,\alpha,t).
\end{equation*}

\noindent
Thus

% ---------------------------------
\begin{equation*}
\int_0^{2\pi} \int_0^{\infty} f d \rho d \theta = g(\omega) h(\alpha).
\end{equation*}

\noindent
By conservation of the number of oscillators and Eqs. (\ref{eq:LanStuCouP}),
$f$ satisfies the following continuity equation,

% --------------------------------------------------------------- f_evol
\begin{align}
\begin{split}
\frac{\partial f}{\partial t} 
&+ \frac{\partial}{\partial \rho} \left\{ \left[
  \alpha \rho - \rho^3 + \frac{\Gamma}{2} \left( e^{-i\theta} \langle z \rangle
                     + e^{i\theta} \langle z \rangle^* \right) \right]f \right\} \\
&+ \frac{\partial}{\partial \theta} \left\{ \left[
  \omega - \bar{\gamma}\rho^2 + \frac{\Gamma}{2i\rho} 
                     \left( e^{-i\theta} \langle z \rangle
                     - e^{i\theta} \langle z \rangle^* \right) \right]f \right\} = 0,
\end{split}
\label{eq:f_evol}
\end{align}

\noindent
where
% ---------------------------------------------------------------- ordp2
\begin{equation}
\langle z \rangle = \iiint \rho e^{i\theta} f d\rho d\theta d\omega d\alpha.
\label{eq:ordp2}
\end{equation}

\noindent
For $\alpha > 0$ the time independent incoherent ($\langle z \rangle_+ \equiv 0$)
solution of (\ref{eq:f_evol}) and (\ref{eq:ordp2}) is 
% ----------------------------------------------------------------- den_sol
\begin{equation}
f_0 = \frac{g(\omega)h(\alpha)}{2\pi} \delta(\rho-\sqrt{\alpha}).
\label{eq:den_sol}
\end{equation}

\noindent
We now introduce a perturbation to the solution (\ref{eq:den_sol}),
% ---------------------------------------------------------------- den_pert
\begin{equation}
f = f_0 + e^{st-i\theta}\delta f + \{ O.P.T.\},
\label{eq:den_pert}
\end{equation}

\noindent
where $\{ O.P.T.\}$ denotes ``other perturbation terms'' whose
$\theta$ variation is proportional to $\exp(i n \theta)$ with
$n \neq -1$. These other terms do not contribute to $\langle z \rangle_+$
[see Eq. (\ref{eq:ordp2})] and so turn out to be of no consequence to what follows.
Inserting (\ref{eq:den_pert}) and (\ref{eq:den_sol}) into (\ref{eq:f_evol})
we obtain for $\alpha > 0$

% ---------------------------------------------------------------- denp_evol
\begin{equation}
\begin{split}
(s-i\omega&+i\bar{\gamma}\rho^2) \delta f + \frac{\partial}{\partial \rho}
\left[ (\alpha-\rho^2) \rho \delta f\right] = \\
&\frac{\Gamma \langle \delta z \rangle g(\omega) h(\alpha)}{4\pi} \left\{
\frac{\delta(\rho-\sqrt{\alpha})}{\sqrt{\alpha}} - \delta'(\rho-\sqrt{\alpha})
\right\},
\end{split}
\label{eq:denp_evol}
\end{equation}

\noindent
where
% ----------------------------
\begin{equation*}
\delta'(\rho-\sqrt{\alpha}) = \frac{d}{d\rho} \delta(\rho-\sqrt{\alpha}).
\end{equation*}

{\it Calculation of $\langle \delta z \rangle_+$}. We now solve
(\ref{eq:denp_evol}) for $\delta f$. To do this we assume a solution
of the form

% -------------------------------------------------------------- denp_evol_antz
\begin{equation}
\delta f = \frac{\Gamma \langle \delta z \rangle g(\omega) h(\alpha)}{4\pi}
[c_1(\omega,s) \delta(\rho-\sqrt{\alpha}) + c_2(\omega,s) \delta'(\rho-\sqrt{\alpha})],
\label{eq:denp_evol_antz}
\end{equation}

\noindent
and substitute this assumed form into (\ref{eq:denp_evol}). Using the delta 
function identities 

% ------------------------------------------------------------ 
\begin{align*}
F(\rho)\delta(\rho-\sqrt{\alpha})  &= F(\sqrt{\alpha})\delta(\rho-\sqrt{\alpha}), \\
F(\rho)\delta'(\rho-\sqrt{\alpha}) &= F(\sqrt{\alpha})\delta'(\rho-\sqrt{\alpha})
                               - F'(\sqrt{\alpha})\delta(\rho-\sqrt{\alpha}),
\end{align*}

\noindent
(where the second of these identities follows from differentiating the first), 
Eq. (\ref{eq:denp_evol}) yields

% ---------------------------------------------------------------- procd1
\begin{equation}
T_1 + T_2 = \frac{1}{\sqrt{\alpha}} \delta - \delta'
\label{eq:procd1}
\end{equation}

\noindent
where $\delta = \delta(\rho-\sqrt{\alpha})$, $\delta'=\delta'(\rho-\sqrt{\alpha})$,
$T_1$ results from the first term on the left hand side of (\ref{eq:denp_evol}),

% -------------------------
\begin{eqnarray*}
T_1 &= (s-i\omega&+i\bar{\gamma}\rho^2) (c_1\delta+c_2\delta') \\
    &= (s-i\omega&+i\bar{\gamma}\alpha) (c_1\delta+c_2\delta') 
       -2i \bar{\gamma} \sqrt{\alpha} c_2 \delta,
\end{eqnarray*}

\noindent
and $T_2$ results from the second term on the left hand side of 
(\ref{eq:denp_evol}),

% -------------------------
\begin{equation*}
T_2 = \frac{\partial}{\partial \rho} \{ (\alpha-\rho^2) 
\rho (c_1 \delta + c_2 \delta') \} = 2 \alpha c_2 \delta'.
\end{equation*}

Separately equating coefficients of $\delta$ and $\delta'$
on the two sides of (\ref{eq:procd1}), we obtain two linear equations
for the coefficients $c_1$ and $c_2$. Solution of these equations
yields 

% -------------------------
\begin{align*}
c_1 &= \frac{1}{\sqrt{\alpha}} \frac{s-i\omega+2\alpha-i\bar{\gamma}\alpha}
                                   {s-i\omega+2\alpha+i\bar{\gamma}\alpha}
                         \centerdot \frac{1}{s-i\omega+i\bar{\gamma}\alpha}, \\
c_2 &= -\frac{1}{s-i\omega+i\bar{\gamma}\alpha+2\alpha}. 
\end{align*}

\noindent
Insertion of (\ref{eq:denp_evol_antz}) with these expressions 
for $c_1$ and $c_2$ into (\ref{eq:ordp2}) then yields $\langle \delta z \rangle_+$,

% ------------------------------------------------------------------------ delz+
\begin{equation}
\begin{split}
\langle \delta z \rangle_+ = &\Gamma \langle \delta z \rangle
\int_{-\infty}^{+\infty} d\omega g(\omega) \\ 
&\left\{
\int_0^{+\infty} \frac{(s-i\omega+\alpha) h(\alpha)}
{[s-i\omega+2\alpha+i\bar{\gamma}\alpha][s-i\omega+i\bar{\gamma}\alpha]} d\alpha
\right\}.
\end{split}
\label{eq:delz+}
\end{equation}

\noindent
Note that the $\alpha$ integration in (\ref{eq:delz+}) is only over 
positive $\alpha$ (i.e., the integration runs from $\alpha=0$ to 
$\alpha=\infty$.)

{\it Equation determining s}. Inserting (\ref{eq:incohN_pert_avg}) and (\ref{eq:delz+})
into (\ref{eq:ordp_pert1}) we obtain,

% ----------------------------------------------------------------- Gaminv
\begin{equation}
\begin{split}
\Gamma^{-1} = &\int_{-\infty}^{\infty} \int_0^{+\infty}
       \frac{(s-i\omega+\alpha) g(\omega)h(\alpha)d\alpha d\omega}
       {[s-i\omega+2\alpha+i\bar{\gamma}\alpha][s-i\omega+i\bar{\gamma}\alpha]} \\
&+ \int_{-\infty}^{\infty} \int_{-\infty}^{0}
    \frac{g(\omega)h(\alpha)d\alpha d\omega}{s+|\alpha|-i\omega} \equiv D(s). 
\end{split}
\label{eq:Gaminv}
\end{equation}

\noindent
By causality, this expression for the dispersion function $D(s)$, as well
as our previous results, Eqs. (\ref{eq:incohN_pert_avg}) and (\ref{eq:delz+}),
for $\langle \delta z \rangle_-$ and $\langle \delta z \rangle_+$,
are defined with $Re(s)>0$. This implies the $\omega$-integration contour
should pass above all poles in the complex $\omega$-plane. We note
that for $Re(s)>0$ the $\omega$-integration poles in (\ref{eq:incohN_pert_avg}), (\ref{eq:delz+})
and (\ref{eq:Gaminv}) all lie in the lower half $\omega$-plane.
Since we are interested in the occurrence of instability,
and instability corresponds to $Re(s)>0$, the form 
giving $D(s)$ by (\ref{eq:Gaminv}) is sufficient for our purposes
($D(s)$ for $Re(s) \leq 0$ can be obtained by analytic continuation, from 
the $Re(s)>0$ result). 

% ==================================================================== Lorent_freq_dist
\section{Lorentzian Frequency Distribution} \label{sec:Lorent_freq_dist}
As discussed in Sec. \ref{sec:intro}, and as we will verify by the example in 
Sec. \ref{sec:freqdist_effect}, we believe that
different monotonically decreasing, continuous frequency distribution
functions $g(\omega)$ often (but not always Ref. \cite{Omel'chenko})
yield similar qualitative behavior,
and we, therefore,
specialize here to one such $g(\omega)$ that allows
easy analytic evaluation
of the integrals over $\omega$, namely, the case of Lorentzian $g(\omega)$,
% ----------------------------------------------------------- Lorent
\begin{equation}
g(\omega) = \frac{1}{\pi} \frac{1}{\omega^2+1} = \frac{1}{2 \pi i} \left\{
\frac{1}{\omega-i} - \frac{1}{\omega+i} \right\},
\label{eq:Lorent}
\end{equation}

\noindent
where we have adopted a normalization of $t, \Gamma$ and 
$\alpha$ so that the half-width of
$g(\omega)$ is one ($g(0)=2g(1)$). Since $Re(s)>0$, the only 
$\omega$-pole of the integrands in (\ref{eq:Gaminv}) that is 
located in $Im(\omega) \geq 0$ is the one at $\omega=i$ 
[see Eq. (\ref{eq:Lorent})]. In addition, the magnitudes of the integrands behave
like $|\omega|^{-3}$ for large $|\omega|$. Thus, we can deform the $\omega$-integration
path by shifting it upward into the complex $\omega$-plane, letting $Im(\omega)$
along the path approach $+\infty$. The integration then yields the residue of the pole
at $\omega=i$,
% --------------------------------------------------------- Disp
\begin{equation}
\begin{split}
D(s) = &\int_0^{+\infty} \frac{(s+1+\alpha) h(\alpha) d\alpha }
   {[s+1+2\alpha+i\bar{\gamma}\alpha][s+1+i\bar{\gamma}\alpha]} \\
 &+ \int_{-\infty}^0 \frac{h(\alpha) d\alpha}{s+|\alpha|+1}. 
\end{split}
\label{eq:Disp}
\end{equation}

%\noindent
In Sec. \ref{sec:condlinsta} we investigate conditions under which
Eq. (\ref{eq:Disp}) predicts instability (i.e., existence of a solution to
$D(s)=\Gamma^{-1}$ with $Re(s)>0$).

% =========================================================== condlinsta (alp_spread)
\section{Condition for Instability of the $\langle z \rangle =0$ state: 
The effect of a spread in the growth
rates in $\alpha$} \label{sec:condlinsta}
In this section, we consider the case where there
is no nonlinear frequency shift (i.e., $\bar{\gamma}=0$), with 
$g(\omega)$ being Lorentzian. 
Using a generalization of the technique in Ref. \cite{Mirollo}
(see proof of their Theorem 2), it can be shown that the solution of
$D(s)=1/\Gamma$ is real. Thus, as we pass from stability to instability,
$s$ goes through $s=0$. This results in the following general condition
for instability,
% ------------------------------------------------------- Cond_insta
\begin{equation}
\Gamma > \frac{1}{D(0)},
\label{eq:Cond_insta}
\end{equation}

\noindent
and (\ref{eq:Disp}) and (\ref{eq:Cond_insta}) imply that instability occurs when $\Gamma$ exceeds 
the critical value $\Gamma_c$ given by

% ------------------------------------------------------------- Gam_c1
\begin{equation}
\Gamma_c = \left\{ \int_0^{\infty} \frac{\alpha+1}{2\alpha+1} h(\alpha) d\alpha
+ \int_{-\infty}^0 \frac{1}{1-\alpha}h(\alpha) \right\}^{-1}.
\label{eq:Gam_c1}
\end{equation}

As a simple reference case, we first consider (\ref{eq:Gam_c1}) 
when there is no dispersion in $\alpha$,
% ----------------------------
\begin{equation*}
h(\alpha) = \delta (\alpha - \bar{\alpha})
\end{equation*}

\noindent
in which case we obtain
% -------------------------------------------------------------------- Gam_c2
\begin{equation}
\Gamma_c = \left\{
\begin{array}{l l}
(2\bar{\alpha}+1)/(\bar{\alpha}+1), 
            &\quad  \text{for $\bar{\alpha} \geq 0$}, \\
1+|\bar{\alpha}|, 
            &\quad  \text{for $\bar{\alpha} \leq 0$}.
\end{array} \right.
\label{eq:Gam_c2}
\end{equation}

\begin{figure}
  {\includegraphics[width=10cm]{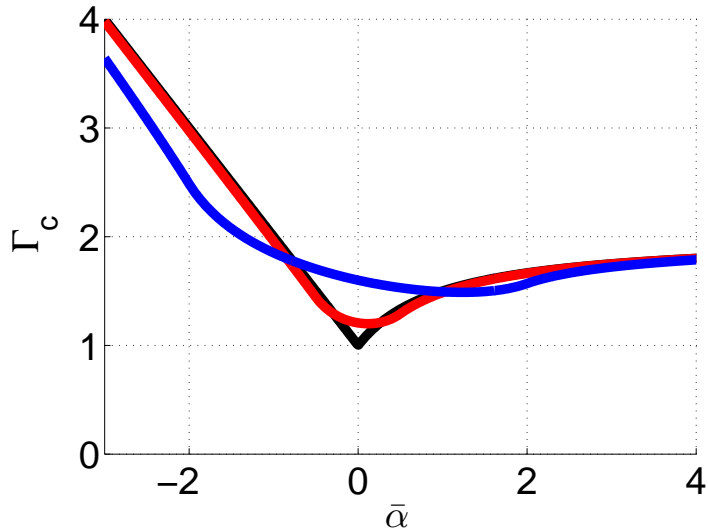}} 
  \caption{(Color online) The effect of dispersion in $\alpha$ ($\bar{\gamma}=0$) with
 a Lorentzian $g(\omega)$.
Stability/Instability regions of $\Gamma-\bar{\alpha}$ space for several 
different values of spread  $\delta{\alpha}$ in 
the linear growth parameter $\alpha$ with mean $\bar{\alpha}$ 
[Legend: Black line ($\delta{\alpha} = 0$), red line ($\delta{\alpha} = 0.5$),
blue line ($\delta{\alpha} = 2$)].}
  \label{fig:Gamc_delalp_gLor}
\end{figure}

\noindent
The resulting phase diagram is given by the black line 
in Fig. \ref{fig:Gamc_delalp_gLor}.
This result (with the different parametrization used in Eq. (\ref{eq:LanStu_oldw}))
has been previously obtained in Refs. \cite{Matthews,Matthews2}.
Note that $\Gamma_c \rightarrow 2$ as $\bar{\alpha} \rightarrow + \infty$.
The value $\Gamma_c=2$ is the critical coupling value that
applies for the Kuramoto model with a Lorentzian frequency pdf,
Eq. (\ref{eq:Lorent}). The applicability of the Kuramoto result for
large $\bar{\alpha}$ can be understood from Eq. (\ref{eq:LanStuCouP_rho})
with $\Gamma$ neglected,  $d\rho/dt = \bar{\alpha}\rho-\rho^3$,
which when linearized about the incoherent equilibrium value, 
$\rho = \sqrt{\bar{\alpha}}$, yields $d\delta\rho/dt=-2\bar{\alpha}\delta\rho$.
Thus perturbations from $\rho = \sqrt{\bar{\alpha}}$ relax at the exponential
rate $2 \bar{\alpha}$, and, for large $\bar{\alpha}$, this rate
becomes much faster than the other relevant time scale, namely,
the spread in $\omega$ (which we have normalized to $1$). Hence, 
for $\bar{\alpha} \gg 1$, the oscillator amplitude is essentially
frozen, and the Kuramoto oscillator description is valid. As shown in
Fig. \ref{fig:Gamc_delalp_gLor} and Eq. (\ref{eq:Gam_c2}), when $\bar{\alpha} \gg 1$
does not hold, the effect of amplitude dynamics is to reduce 
$\Gamma_c$ (for $\bar{\alpha} \geq 0$) from the Kuramoto value 
with the reduction increasing to a factor of $2$ as 
$\bar{\alpha} \rightarrow 0^+$ ($\Gamma_c=2$ at $\bar{\alpha} \rightarrow + \infty$
in comparison with $\Gamma_c=1$ at $\bar{\alpha}=0$). An additional
interesting point is that comparison of the black line in
Fig. \ref{fig:Gamc_delalp_gLor} with the
results for the phase diagram in the case of Belousov-Zhabotinsky system
of Ref. \cite{Taylor} (see discussion in Sect. \ref{sec:intro}) shows
a striking qualitative similarity between the two 
(e.g., see Fig. 3 of Ref. \cite{Taylor}).

Referring to Eq. (\ref{eq:Gam_c2}) and Fig. \ref{fig:Gamc_delalp_gLor}, we see that
there is a sharp transition in behavior as $\bar{\alpha}$ crosses
$\bar{\alpha}=0$. In particular, the $\langle z \rangle = 0$ state for 
$\bar{\alpha}<0$ results from the fact that $z_i=0$ for all oscillators, while for
$\bar{\alpha}>0$ all oscillators have $|z_i|=\sqrt{\bar{\alpha}} > 0$
and $\langle z \rangle = 0$ results from incoherence of the individual 
oscillator phases. This sharp transition in behavior is reflected by 
the discontinuity of the derivative, $d\Gamma_c/d\bar{\alpha}$, at 
$\bar{\alpha}=0$. The sharp nature of the transition at 
$\bar{\alpha}=0$ is, however, a nonphysical artifact of the assumption
of no dispersion in the individual oscillator growth / damping rates
used in obtaining (\ref{eq:Gam_c2}). In typical physical situations,
such as the experiment in Ref. \cite{Taylor} (see discussion in 
Sect. \ref{sec:formulation}), dispersion in $\alpha$ is to be expected
(Fig. \ref{fig:sch_h}). To simply illustrate its effect we consider
the example where $h(\alpha)$ is uniform within some range $\delta \alpha$
about an average value $\bar{\alpha}$,

% ----------------------------------------------------------- h_spread
\begin{equation}
h(\alpha) = (2 \delta \alpha)^{-1} U(\delta \alpha - |\alpha-\bar{\alpha}|),
\label{eq:h_spread}
\end{equation}

\noindent
where $U(x)$ denotes the unit step function; $U(x)=1$ for $x \geq 0$ and
$U(x)=0$ for $x<0$. 
Using (\ref{eq:h_spread}) in (\ref{eq:Gam_c1}),
we get 
for $\bar{\alpha} > \delta \alpha$,
% ------------------------------------------------------------------------------------ Gam_c3_1
\begin{equation}
\Gamma_c^{-1} = \frac{1}{2 \delta \alpha} \left[\delta \alpha + \frac{1}{4}\ln \left( 
       \frac{\bar{\alpha}+\delta \alpha + 1/2}{\bar{\alpha}-\delta \alpha + 1/2}
        \right) \right]; 
\label{eq:Gam_c3_1} 
\end{equation}

\noindent
for  $\bar{\alpha} <- \delta \alpha$,
% ----------------------------------------------------------------------------- Gam_c3_2
\begin{equation}
\Gamma_c^{-1} = \frac{1}{2 \delta \alpha} \ln \left( 
       \frac{1-\bar{\alpha}+\delta \alpha}{1-\bar{\alpha}-\delta \alpha}
        \right); 
       \label{eq:Gam_c3_2}
\end{equation}

\noindent
and for $|\bar{\alpha}| < \delta \alpha$,
% -------------------------------------------------------------------------------- Gam_c3_3
\begin{equation}
\Gamma_c^{-1} = \frac{1}{2 \delta \alpha} \left[ \frac{\bar{\alpha}+\delta \alpha}{2}
      + \ln(1-\bar{\alpha}+\delta \alpha) + \frac{1}{4}\ln(1+2\bar{\alpha}+
      2\delta \alpha) \right]. 
       \label{eq:Gam_c3_3} 
\end{equation}

%\label{eq:Gam_c3}
%\end{subequations}

\noindent
As the dispersion in $\alpha$, $\delta \alpha$, approaches zero,
(\ref{eq:h_spread}) becomes a delta function, and Eqs. (\ref{eq:Gam_c3_1})-(\ref{eq:Gam_c3_3})
reduce to (\ref{eq:Gam_c2}). The other two lines in
Fig. \ref{fig:Gamc_delalp_gLor} show the phase
diagram from Eqs. (\ref{eq:Gam_c3_1})-(\ref{eq:Gam_c3_3})
for two more values of $\delta \alpha$. For $\delta \alpha > 0$
the discontinuity in $d \Gamma_c/d\bar{\alpha}$ (which occurs 
for $\delta \alpha=0$ at $\bar{\alpha}=0$)
is removed by dispersion in $\alpha$, and
the sharp transition that occurs at $\bar{\alpha}=0$ 
(black line in Fig. \ref{fig:Gamc_delalp_gLor})
is now smoothed out \cite{note3}.
Further, it is also noticed that the minimum of $\Gamma_c$ rises and shifts
from $\bar{\alpha}=0$ when $\delta \alpha = 0$
to $\bar{\alpha} > 0$ when $\delta \alpha > 0$.

% ffffffffffffffffffffffffffffffffffffffffffffff

% ================================================================= nonlinear_effect
\section{Condition for instability of the $\langle z \rangle =0$ state:
The effect of a nonlinear frequency shift}\label{sec:nonlinear_effect}
We now address the effect of nonlinear frequency shift, $\bar{\gamma} \neq 0$, 
and we consider the simple case of no dispersion in $\alpha$, 
$h(\alpha)=\delta(\alpha - \bar{\alpha})$ again for the
case of Lorentzian $g(\omega)$. We note from Eq. (\ref{eq:LanStuCou}), if the distribution 
of $\omega_i$ values is symmetric, then positive and negative values of 
$\bar{\gamma}$ are equivalent ($z_i \rightarrow z_i^*$). Thus, we consider
$\bar{\gamma} > 0$ only.
As is evident
from Eq. (\ref{eq:Disp}), $\bar{\gamma}$ has no effect on the linear theory
for $\bar{\alpha}<0$, and, consequently, the result for $\Gamma_c$ given
by Eq. (\ref{eq:Gam_c2})
still applies for $\bar{\alpha} \leq 0$. For 
$\bar{\alpha}>0$, however, the effect of a nonlinear frequency shift
can be substantial. Equation (\ref{eq:Disp}) for 
$h(\alpha)=\delta(\alpha-\bar{\alpha})$, $\bar{\alpha}>0$ gives

% ----------------------------------------------------------- Disp2
\begin{equation} 
\Gamma^{-1} = D(s) = \frac{s+1+\bar{\alpha}}
     {(s+1+2\bar{\alpha}+i\bar{\gamma}\bar{\alpha})(s+1+i\bar{\gamma}\bar{\alpha})},
\label{eq:Disp2}
\end{equation}

\noindent
which yields a quadratic equation for $s$, solution of which can be used to
obtain stability boundary curves in $\Gamma-\bar{\alpha}$ space. At the transition
point, $Re(s)$ goes through zero. Substituting $s=i \Omega$ into
Eq. (\ref{eq:Disp2}) and separating the real and imaginary parts, $\Gamma_c$
and $\Omega$ are then given by the solution of the following pair of equations
% ------------------------------------------------------------ Gam_NLc
\begin{subequations}
\begin{align}
1-\Omega^2+2 \bar{\alpha}(1-\bar{\gamma}\Omega) - \bar{\alpha}^2 \bar{\gamma}^2
& = \Gamma_c (1+\bar{\alpha}), \label{eq:Gam_NLc1} \\
2(1+\bar{\alpha})(\Omega+\bar{\alpha}\bar{\gamma}) &= \Gamma_c \Omega. \label{eq:Gam_NLc2}
\end{align}
\label{eq:Gam_NLc}
\end{subequations}

\begin{figure}
  {\includegraphics[width=10cm,height=8cm]
              {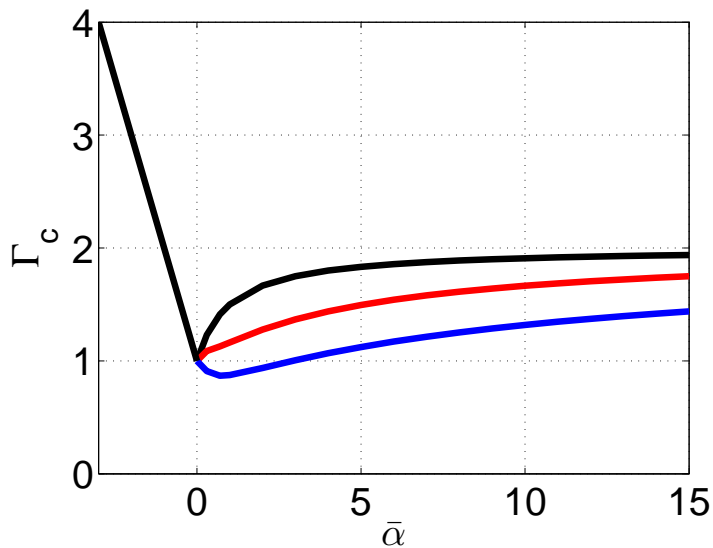}} 
  \caption{(Color online) Stability / Instability regions of $\Gamma-\bar{\alpha}$ space for several 
different values of the nonlinear frequency shift parameter $\bar{\gamma}$
[Legend: Black line ($\bar{\gamma} = 0$), red line ($\bar{\gamma} = 2$),
blue line ($\bar{\gamma} = 4$)]. Notice that the three lines
coincide when $\bar{\alpha}<0$.}
\label{fig:Gamc_bgamma_gLor}
\end{figure}

\noindent
When $\bar{\gamma}=0$ (note $\Omega=0$ for this case), 
the solution for the critical coupling strength
of (\ref{eq:Gam_NLc}) is given by
% --------------------------------------------------------- Gam_NLc0
\begin{equation}
\Gamma_{c0} = \frac{1+2\bar{\alpha}}{1+\bar{\alpha}},
\label{eq:Gam_NLc0}
\end{equation}

\noindent
by which
(\ref{eq:Gam_NLc1}) can be rearranged to give
% ------------------------------------------------------------- Gam_NLc3
\begin{equation}
\Gamma_c = \Gamma_{c0} - \frac{(\Omega+\bar{\alpha}\bar{\gamma})^2}{1+\bar{\alpha}},
\label{eq:Gam_NLc3}
\end{equation}

\noindent
which shows that the effect of $\bar{\gamma}$ is always to decrease 
$\Gamma_c$. Figure \ref{fig:Gamc_bgamma_gLor} shows the values of
$\Gamma_c$ as a function of $\bar{\alpha}$ for several different values
of $\bar{\gamma}$ ($\bar{\gamma}=0$ plotted in black,
$\bar{\gamma}=2$ plotted in red, and $\bar{\gamma}=4$ plotted in blue). 
By solving for $\Omega$ in (\ref{eq:Gam_NLc2})
and substituting it back in (\ref{eq:Gam_NLc1}), we obtain
% ------------------------------------------------------------- Gam_NLc4
\begin{equation}
\Gamma_c = \Gamma_{c0}-\frac{1}{1+\bar{\alpha}} 
\left[ \frac{\bar{\alpha}\bar{\gamma}\Gamma_c}{\Gamma_c-2(1+\bar{\alpha})} \right]^2.
\label{eq:Gam_NLc4}
\end{equation}

\noindent
Equation (\ref{eq:Gam_NLc4}) shows that
$\Gamma_c \rightarrow 2$ as $\bar{\alpha} \rightarrow +\infty$. As seen
in Fig. \ref{fig:Gamc_bgamma_gLor}, increasing $\bar{\gamma}$ eventually 
moves the minimum of $\Gamma_c$ below one and shifts the location of 
the minimum into $\bar{\alpha}>0$.

% fffffffffffffffffffffffffffffffffffffffffffffffffff

% ================================================================ freqdist_effect
\section{The effect of the frequency distribution function} \label{sec:freqdist_effect} %\label{sec:numerical}
In Secs. \ref{sec:condlinsta} and \ref{sec:nonlinear_effect} we consider the effect of a spread
in $\alpha$ and of a nonlinear frequency shift for the illustrative case of a Lorentzian 
distribution function of the oscillator natural frequencies, Eq. (\ref{eq:Lorent}). We now
ask how might these results be altered if a different frequency distribution were used. We note
that the Lorentzian decays rather slowly for large $\omega$, $g(\omega) \sim \omega^{-2}$.
Thus to test dependence on the form of $g(\omega)$, we will examine another distribution function which
is very different from the Lorentzian, in that it has a sharp cutoff to $g(\omega)=0$ as $\omega$
increases. In particular, we will consider a ``flat-top" distribution, that is uniform 
in $-1 \leq \omega \leq 1$ and zero otherwise, 
% ------------------------------------------------------- Uniform
\begin{equation}
g(\omega) = \frac{1}{2} U(1 - |\omega|),
\label{eq:Uniform}
\end{equation}

\noindent
where $U(x)$ is the unit step function.
In spite the qualitatively different large $|\omega|$ behavior of the Lorentzian and the 
flat-top $g(\omega)$ distributions, we will find that the resulting stability conditions show
qualitatively similar behavior.

The calculation of $\Gamma_c$ with $g(\omega)$ given by (\ref{eq:Uniform}) 
is done by using (\ref{eq:Gaminv}) (see the Appendix).
In Fig. \ref{fig:Gamc_delalp_gUnif} we show the dependence of $\Gamma_c$ 
on $\bar{\alpha}$ for several different values of $\delta \alpha$,
where $h(\alpha)$ is given by (\ref{eq:h_spread}) and
$\bar{\gamma}=0$
for all oscillators. A comparison between
Fig. \ref{fig:Gamc_delalp_gUnif} and Fig. \ref{fig:Gamc_delalp_gLor}
reveals remarkably similar dependence,
apart from a difference in the vertical scale due to different functional 
dependence of $g(\omega)$ \cite{note4}.
Next, we consider the dependence of $\Gamma_c$ on
$\bar{\gamma}$ when $h(\alpha)=\delta (\alpha - \bar{\alpha})$
with $g(\omega)$ given by (\ref{eq:Uniform}). 
Figure \ref{fig:Gamc_bgamma_gUnif} shows 
the dependence of $\Gamma_c$ on $\bar{\alpha}$
for several different values of $\bar{\gamma}$.
The black line shows the result
when $\bar{\gamma}=0$, which is the same black line 
in Fig. \ref{fig:Gamc_delalp_gUnif}. The other two lines are obtained
by numerically  solving Eq. (\ref{eq:gUnif_NL_Gamc1}) in the Appendix when
$\bar{\gamma} \neq 0$.  In comparison with Fig. \ref{fig:Gamc_bgamma_gLor}, 
we see similar dependence in that, as $\bar{\gamma}$ increases, $\Gamma_c$ decreases.

% ffffffffffffffffffffffffffffffffffffffffffffffffffffffffffffffffffffffffffffffff
\begin{figure}
  {\includegraphics[width=10cm]{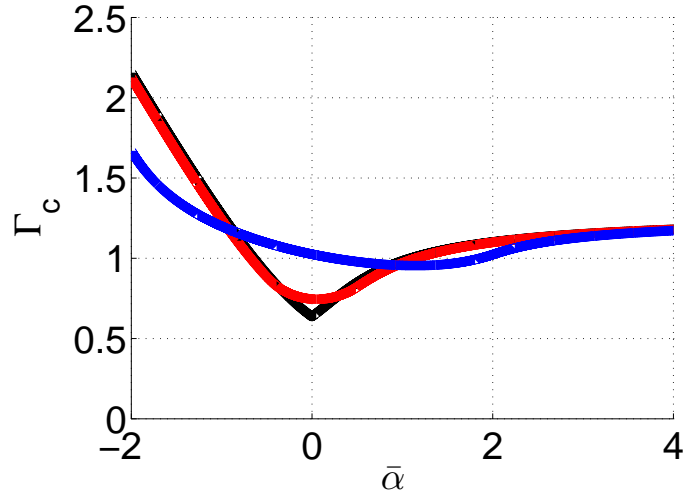}} 
  \caption{(Color online) The effect of dispersion in $\alpha$ with a 
uniformly distributed $g(\omega)$, Eq. (\ref{eq:Uniform}).
Stability/Instability regions of $\Gamma-\bar{\alpha}$ space for several 
different values of spread  $\delta{\alpha}$ in 
the linear growth parameter $\alpha$ with mean $\bar{\alpha}$ 
[Legend: Black line ($\delta \alpha = 0$), red line ($\delta \alpha = 0.5$),
blue line ($\delta \alpha = 2.0$)].}
  \label{fig:Gamc_delalp_gUnif}
\end{figure}

\begin{figure}
  {\includegraphics[width=10cm,height=8cm]
              {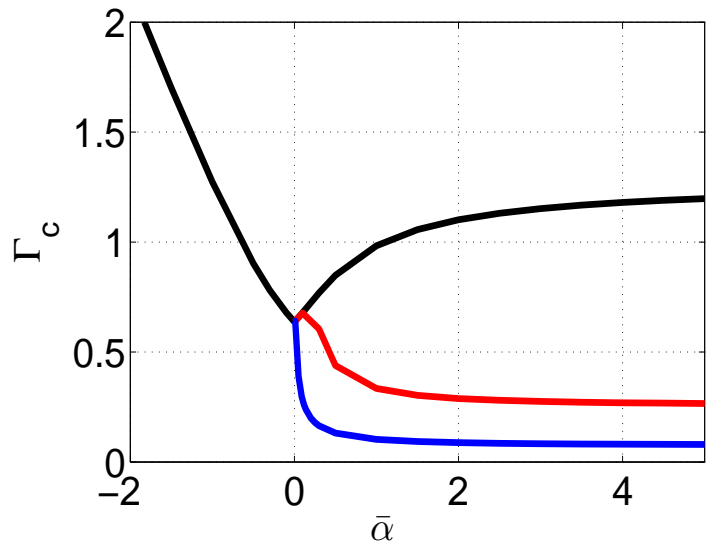}} 
  \caption{(Color online) Stability / Instability regions of $\Gamma-\bar{\alpha}$ space for several 
different values of the nonlinear frequency shift parameter $\bar{\gamma}$
[Legend: Black line ($\bar{\gamma} = 0$), red line ($\bar{\gamma} = 2$),
blue line ($\bar{\gamma} = 4$)]. Notice that the three lines
coincide when $\bar{\alpha}<0$.}
\label{fig:Gamc_bgamma_gUnif}
\end{figure}

% ============================================================== above_instab
\section{Nonlinear phenomena above the instability threshold with finite $\alpha$-spread
and nonlinear frequency shift}  \label{sec:above_instab}
In the previous sections, we calculated the critical coupling strength $\Gamma_c$
marking the onset of instability of the quiescent state
($\langle z \rangle = 0$). 
Above the critical value $\Gamma_c$, we find that
Landau-Stuart oscillator
networks exhibit a rich variety of collective behavior. We now briefly
review past work on the nonlinear behavior found above the instability 
threshold.

Matthews et al. \cite{Matthews2} studied nonlinear collective behavior
in the special case that
$\alpha_j = 1$ and $\gamma_j=0$ for all oscillators $(j=1,2, \cdots N)$, and
$g(\omega)$ takes on several different functional forms. 
An important observation in that paper is that
the system behavior can be quite complicated for a range of $\Gamma$ not too far above
$\Gamma_c$. For example, they found period doubling cascades to
chaos, large amplitude oscillations, quasiperiodicity, and
hysteretic behavior close to the boundaries between
different macroscopic states. However, when $\Gamma$ is sufficiently far from $\Gamma_c$, 
the system was always found to settle into
a steady oscillatory state, $\langle z \rangle = constant \times \exp(i \Omega t)$
for some constant $\Omega$. We refer to this as a ``locked state," which 
we define as a solution of (\ref{eq:f_evol}) and (\ref{eq:ordp2}) for which 
the oscillator distribution $f$ has dependence on $(\rho,\theta,t)$ of the form
$f = f(\rho,\theta-\Omega t)$; i.e., the entire distribution rigidly rotates about the
origin of the complex $z$-plane with the uniform rotation rate $\Omega$.

When the nonlinear frequency shift parameter $\bar{\gamma}$ is nonzero
\cite{Hakim,Nakagawa0,Nakagawa1,Nakagawa2}, the system can 
exhibit additional types of complicated coherent behavior. 
For example, Refs. \cite{Hakim,Nakagawa0,Nakagawa1,Nakagawa2,Daido2} studied
systems closely related to Eq. (\ref{eq:LanStuCou}), but with homogeneous parameters.
An important feature found in those references is the tendency 
for the system to form clusters (a ``cluster" in this case
is defined as a group of oscillators
which behave identically). Further, depending on parameter values and on
initial conditions, the systems can 
form cluster states of varying sizes.
In Refs. \cite{Nakagawa0} and \cite{Nakagawa1}, the authors also found
chaotic behavior.

\begin{figure}
 \begin{center}
  \subfigure[$\Gamma / \Gamma_c = 2$]
  {\includegraphics[width=6cm,height=6cm]            
              {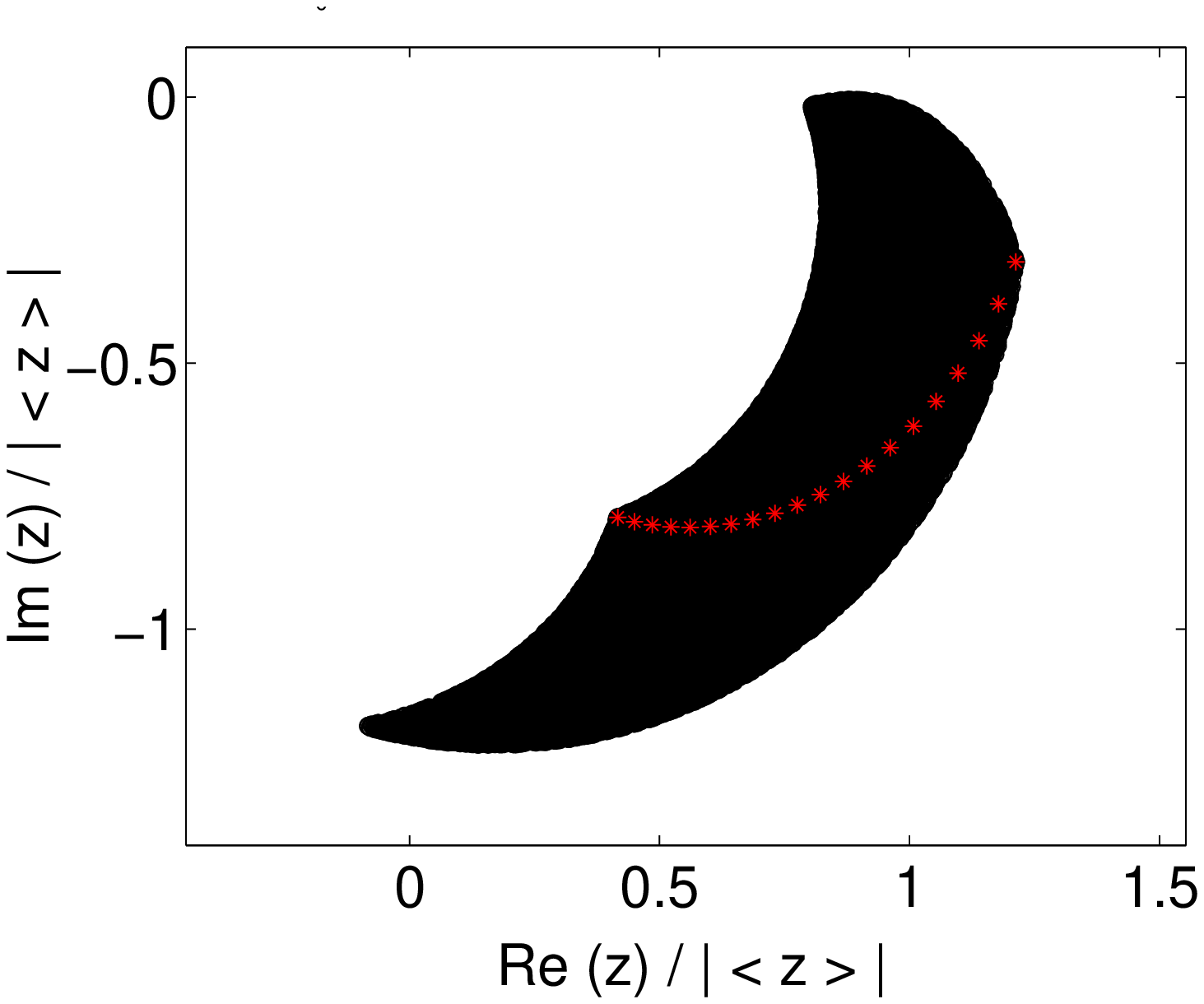} 
                                                       \label{fig:lock_varyGam2}} 
  \subfigure[$\Gamma / \Gamma_c = 10$]
  {\includegraphics[width=6cm,height=6cm]
              {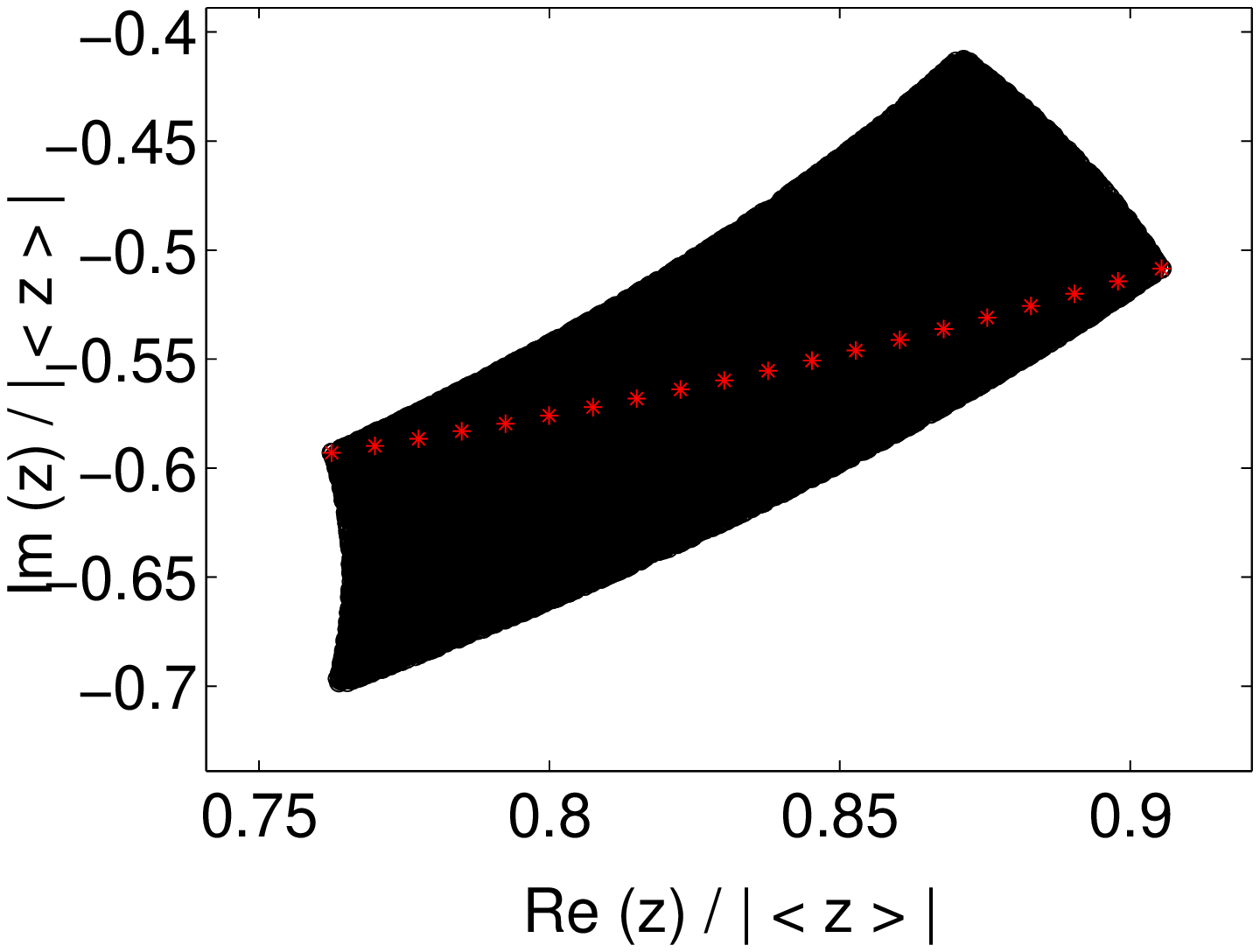} 
                                                       \label{fig:lock_varyGam10}}
 \subfigure[$\Gamma / \Gamma_c = 100$]
  {\includegraphics[width=6cm,height=6cm]
              {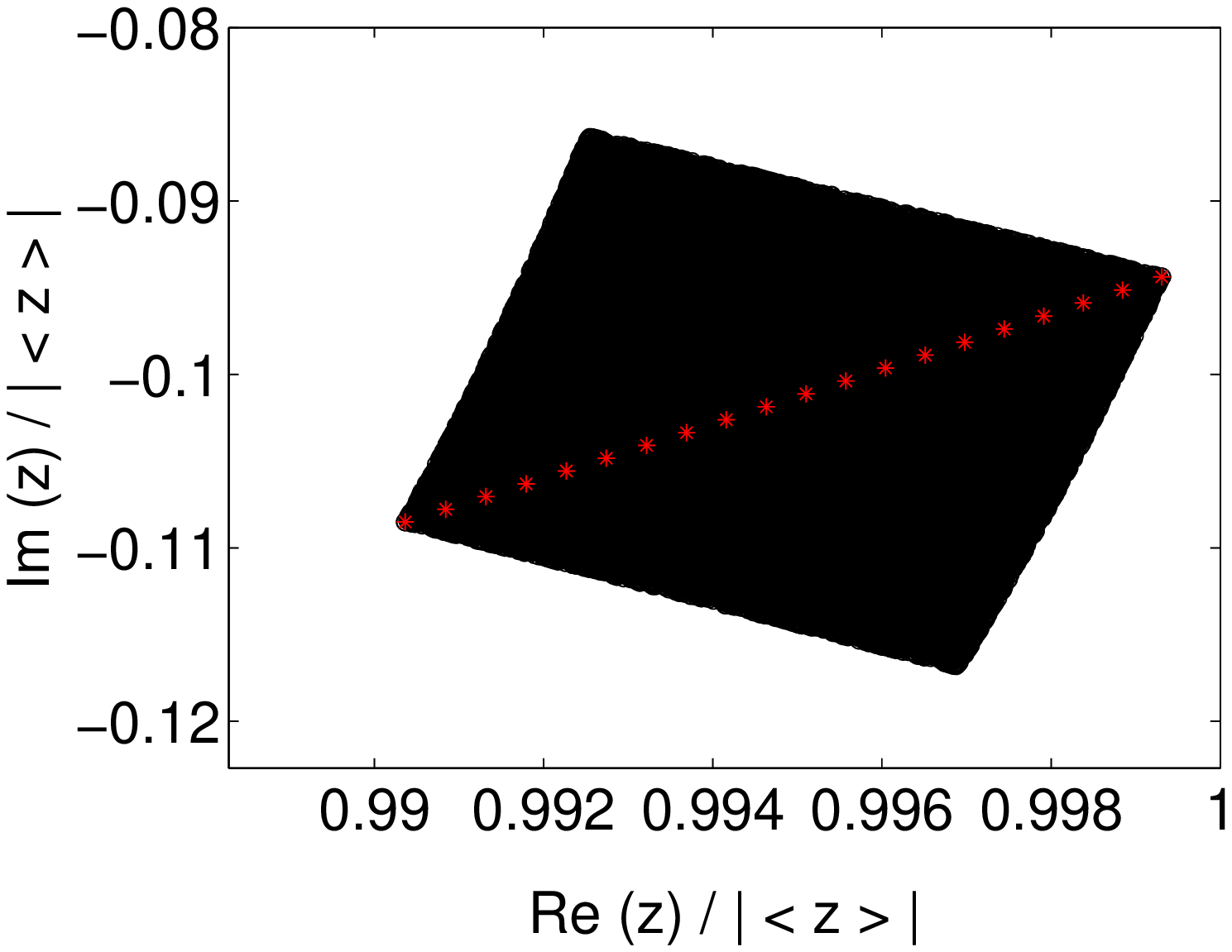} 
                                                       \label{fig:lock_varyGam100}}
 \end{center}
  \caption{(Color online) Locations of $50000$ oscillators (black) in locked states with 
                  different $\Gamma / \Gamma_c$.
                 Twenty oscillators (red cross) of parameter values evenly spaced simultaneously in 
                 $(\alpha,\omega) \in [-0.5,1.5] \times [-1,1]$ are highlighted, i.e., the oscillator with 
                 $(\alpha,\omega)= (-0.5,-1)$ is located at the minimum radius position, and the oscillator with 
                 $(\alpha,\omega)= (1.5,1)$ is located at the maximum radius position, 
                 and other oscillators of intermediate
                 parameter values are distributed in between ($N=50000$, $\bar{\alpha}=0.5$,
                 $\delta \alpha=1.0$, $\bar{\gamma}=0.5$; random initial conditions).
                  }
\label{fig:lock_varyGam}
\end{figure}

We emphasize the finding of Ref. \cite{Matthews2} that, for zero nonlinear frequency shift
$\gamma_i \equiv 0$ the system always goes to a locked state attractor 
when $\Gamma$ is sufficiently large.
Consistent with this, we find that when we 
include spreads in $\alpha$,
and $\omega$, and simultaneously allow
$\beta_j = \bar{\beta} \neq0$ and $\gamma_j = \bar{\gamma} \neq 0$,
it is the case that, as $\Gamma$ is increased, there is always a locked state that
the system may settle into.
Furthermore, we analytically prove that this locked state is the only large $\Gamma$ attractor
(as in \cite{Matthews2} which has $\gamma_i \equiv 0$) {\it provided} that the nonlinear
frequency shift $\bar{\gamma}$ is not too large, but we also find 
that other coexisting attractors may be present
if $\bar{\gamma}$ is large enough. This will be
discussed further in Secs. \ref{sec:LS_strong_couple} and \ref{sec:CS_strong_couple}.
As an example of a locked state, 
Fig. \ref{fig:lock_varyGam} shows snapshots of the long-time asymptotic distributions of
the oscillator $z-$values obtained from numerical simulations of Eq. (\ref{eq:LanStuCou}) with 
$g(\omega)$ given by (\ref{eq:Uniform}), $N=50000$, $h(\alpha)$ given by 
(\ref{eq:h_spread}), $\bar{\alpha}=0.5$, $\delta \alpha = 1.0$, $\bar{\gamma}=0.5$ (corresponding
to $\Gamma_c = 0.89$), for 
successively larger values of $\Gamma / \Gamma_c$, all of which are large enough that a locked
state is achieved (Fig. \ref{fig:lock_varyGam2}: $\Gamma / \Gamma_c=2$, 
Fig. \ref{fig:lock_varyGam10}: $\Gamma / \Gamma_c = 10$, 
Fig. \ref{fig:lock_varyGam100}: $\Gamma / \Gamma_c = 100$). Note that, as
appropriate for a locked state, as time increases, these snapshots rotate uniformly about
the origin at a fixed angular rate $\Omega$.

% fffffffffffffffffffffffffffffffffff

We see in Fig. \ref{fig:lock_varyGam2} that the distribution has finite spreads in both the magnitude
and phase of $z$. Examination of the solution shows that oscillators with smaller (larger) natural 
frequencies $\omega$ tend to occur on the clockwise (counterclockwise)
side of the distribution, while larger (smaller) $\alpha$ tend to occur at larger (smaller)
$|z|$ for fixed argument of $z$. Previous works (e.g., \cite{Matthews2}) did not  consider a distribution of 
$\alpha$ and consequently did not find a spread in $|z|$ at constant argument of $z$ (i.e., the
oscillators are distributed along a curve in the complex $z$-plane). Comparing Figs. \ref{fig:lock_varyGam2},
\ref{fig:lock_varyGam10} and \ref{fig:lock_varyGam100}, we see that the 
spread in $z/|\langle z \rangle|$ becomes 
smaller and smaller as $\Gamma / \Gamma_c$ increases. In fact, we argue 
in Sec. \ref{sec:LS_strong_couple} below that one of the stationary states of this system
is when this
spread goes to zero as $\Gamma / \Gamma_c \rightarrow \infty$ (note the greatly magnified scale
for Fig. \ref{fig:lock_varyGam100}). Note that the oscillators in 
Fig. \ref{fig:lock_varyGam} are contained within a {\it single} region, and 
we subsequently refer to such states as {\it single-cluster}
locked states.

\begin{figure}
   {\includegraphics[width=8cm]
      {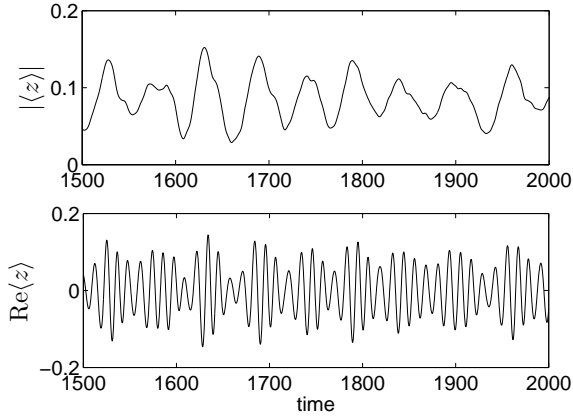}}
   \caption{Time evolution of $|\langle z \rangle|$ (top panel) 
   and $\mbox{Re} \langle z \rangle$ (bottom panel) for a system of 
   $500,000$ oscillators (Parameters: $\bar{\alpha}=0.5$, $\delta \alpha = 1.0$, 
   $\bar{\gamma}=0.5$, $\Gamma / \Gamma_c = 1.07$; random initial conditions.)}
\label{fig:Coh_varyGam}
\end{figure}

% ---------------- added in StuLan9
Figure \ref{fig:Coh_varyGam} illustrates an example of
the occurrence and evolution of a non-locked dynamical 
attractor at lower $\Gamma / \Gamma_c$ with other parameters the same as those in 
Fig. \ref{fig:lock_varyGam}.
In particular, Fig. \ref{fig:Coh_varyGam} shows
$|\langle z \rangle|$ (top panel) and
$\mbox{Re} \langle z \rangle$ (bottom panel)
versus time, after the system has settled into an attractor for $\Gamma / \Gamma_c = 1.07$. 
(Note that for a locked state,  $|\langle z \rangle|$ is constant, 
and $\mbox{Re} \langle z \rangle$
varies sinusoidally in time.)

% fffffffffffffffffffffffffffffffffffffffffffffff

%============================================================ LS_strong_couple
%\section{Destruction of the locked state}
\section{Large coupled Landau-Stuart oscillator networks in the strong coupling limit: 
Single-cluster locked states} 
\label{sec:LS_strong_couple}

In what follows, as in all other previous references (except for 
the weak ``coupling limit'' treatment in Ref. \cite{Montbrio_Pazo1}), we
consider the case where there is no spread in the nonlinear coefficients 
($\gamma_j = \bar{\gamma}$ and $\beta_j = 1$ for all $j$). In this section we ask
why the single-cluster locked state is an attractor for large enough $\Gamma$.

In order to analytically show that a single-cluster
locked state attractor must exist for homogeneous nonlinearity
parameters $\gamma_j \equiv \bar{\gamma}$ and $\beta_j \equiv 1$
at sufficiently large $\Gamma / \Gamma_c$, 
we now consider very large $\Gamma / \Gamma_c$ approximated by taking the limit 
$\Gamma / \Gamma_c \rightarrow \infty$. In particular, using this limit, we will show 
the existence of a simple single-cluster locked 
state and we will demonstrate that it is stable. In Sec. \ref{sec:CS_strong_couple}
we will show that the single-cluster locked state is the
only attractor of the system if $\bar{\gamma}$ is not too large,
but that, when $\bar{\gamma}$ is larger, there can be other coexisting
attractors of various types composed of multiple clusters.

When $\Gamma \gg \alpha_j, \omega_j$ for all $j$, system 
(\ref{eq:LanStuCou}) reduces to 

% --------------------------------------------------------- LanStuNet2
\begin{equation}
\frac{dz_j}{dt} =  - (1 + i\bar{\gamma}) |z_j|^2 z_j+
  \Gamma \langle z \rangle.
\label{eq:LanStuNet2}
\end{equation}

\noindent
Here we have assumed that $|z_j| \gg 1$ in the $\Gamma \rightarrow \infty$ limit. This 
will be subsequently verified.
Alternatively, if the $\omega_j$ are uniform, $\omega_j=\omega$,
and $\Gamma \gg \alpha_j$, even if $\Gamma \gg \omega$ does not
apply, we can still obtain Eq.(\ref{eq:LanStuNet2}) via elimination of
$\omega$ through the transformation $z_j \rightarrow z_j e^{i \omega t}$.
Thus, in this limit, the dynamics is determined by the 
coupling to other oscillators and the nonlinear characteristics of the individual oscillators, 
rather than by the linear properties 
of the individual oscillators.
% ------------------------- added and changed in StuLan8
This is consistent with our numerical tests in Fig. \ref{fig:lock_varyGam}, which
suggests that as $\Gamma / \Gamma_c \rightarrow \infty$, the distribution of oscillators
approaches that of a system of homogeneous parameter values, with the effects of spreads due 
to $\alpha$ and $\omega$ going away (see Fig. \ref{fig:lock_varyGam100}).  
% ----------------------------
We now divide Eq.(\ref{eq:LanStuNet2}) by $\Gamma$, and redefine variables as
% -------------------------------------------------------- LSNet_transfv
\begin{subequations}
\begin{align}
\hat{z}_j &= \frac{z_j}{\sqrt{\Gamma}}, \label{eq:tran_z} \\
\hat{t} &= \Gamma t.  \label{eq:tran_t} 
\end{align}
\label{eq:LSNet_transfv}
\end{subequations}

\noindent
Thus, each term in Eq. (\ref{eq:LanStuNet2}) scales as $\Gamma^{3/2}$ justifying the neglect of the other terms
in Eq. (\ref{eq:LanStuCou}).
Equations (\ref{eq:LanStuNet2}) become
% --------------------------------------------------------- LanStuNet3
\begin{equation}
\frac{d\hat{z}_j}{d\hat{t}} =  - (1 + i\bar{\gamma}) |\hat{z}_j|^2 \hat{z}_j+
\frac{1}{N} \sum_{k=1}^N \hat{z}_k,
\label{eq:LanStuNet3}
\end{equation}

Making the single-cluster 
locked state ansatz $\hat{z}_j = \rho(\hat{t}) \exp [ i \theta(\hat{t}) ]$
gives
% --------------------------------------------------------------- LSCoh
\begin{subequations}
\begin{align}
\frac{d \rho}{d\hat{t}} &= \rho (1- \rho^2), \label{eq:LSCoh_amp} \\
\frac{d \theta}{d\hat{t}} &= - \bar{\gamma} \rho^2. \label{eq:LSCoh_the}
\end{align}
\label{eq:LSCoh}
\end{subequations}

\noindent
This yields the time asymptotic attracting solution,
% --------------------------------------------------------------- LSCoh_asympsol
\begin{equation}
\hat{z_0}(\hat{t}) =  e^{-i \bar{\gamma} \hat{t}}.
\label{eq:LSCoh_asympsol}
\end{equation}

To analyze the stability of (\ref{eq:LSCoh_asympsol}), we perturb
$\hat{z}_0$ to $\hat{z}_0 + e^{-i \bar{\gamma} \hat{t}} 
\delta \hat{z}_j$. From (\ref{eq:LanStuNet3}), the dynamics of 
$\delta \hat{z}_j$ is governed by
% --------------------------------------------------------------- LSCoh_pert
\begin{equation}
\frac{d}{d\hat{t}} \delta \hat{z}_j = \left( -1-i \bar{\gamma} \right)
(\delta \hat{z}_j + \delta \hat{z}_j^*) - \delta \hat{z}_j + \frac{1}{N}
\sum_k \delta \hat{z}_k.
\label{eq:LSCoh_pert}
\end{equation}

\noindent
where ${}^*$ denotes complex conjugation. Similarly, we have
% --------------------------------------------------------------- LSCoh_pertconj
\begin{equation}
\frac{d}{d\hat{t}} \delta \hat{z}_j^* = \left( -1+i \bar{\gamma} \right)
(\delta \hat{z}_j + \delta \hat{z}_j^*) - \delta \hat{z}_j^* + \frac{1}{N}
\sum_k \delta \hat{z}_k^*.
\label{eq:LSCoh_pertconj}
\end{equation}

\noindent
Equations (\ref{eq:LSCoh_pert}) and (\ref{eq:LSCoh_pertconj}) can be regarded
as two independent equations for $\delta \hat{z}_j$ and $\delta \hat{z}_j^*$
respectively.
To study the stability properties of $\delta \hat{z}_j$ and $\delta \hat{z}_j^*$,
consider $\delta \hat{z}_j \sim \delta \hat{Z}_j(s) e^{s \hat{t}}$ and 
 $\delta \hat{z}_j^* \sim \delta \hat{Z}_j^*(s) e^{s \hat{t}}$, for which
Eqs. (\ref{eq:LSCoh_pert}) and (\ref{eq:LSCoh_pertconj}) give

% --------------------------------------------------------------- LSCoh_pertalg
\begin{subequations}
\begin{align}
\left( s+2+i \bar{\gamma} \right) \delta \hat{Z}_j+ 
\left( 1+i \bar{\gamma} \right) \delta \hat{Z}_j^* - 
\langle \delta \hat{Z} \rangle &= 0, \label{eq:LSCoh_pert_Z} \\
\left( s+2-i \bar{\gamma} \right) \delta \hat{Z}_j^*+ 
\left( 1-i \bar{\gamma} \right) \delta \hat{Z}_j - 
\langle \delta \hat{Z}^* \rangle &= 0, \label{eq:LSCoh_pert_Zconj}
\end{align}
\label{eq:LSCoh_pertalg}
\end{subequations}

\noindent
where $\langle \delta \hat{Z} \rangle = N^{-1} \sum_k \delta \hat{Z}_k$,
 $\langle \delta \hat{Z}^* \rangle = N^{-1} \sum_k \delta \hat{Z}_k^*$.
Summing over $j$ we obtain
% ----------------------------------------------------------------- LSCoh_alg2
\begin{equation}
\Upsilon 
\left[ \begin{array}{c}
\langle \delta \hat{Z} \rangle \\
\langle \delta \hat{Z}^* \rangle
\end{array} \right] 
=0,
\label{eq:LSCoh_alg2}
\end{equation}

\noindent
where
% -------------------------------------------------------------- LSCoh_pertmat2
\begin{equation}
\Upsilon \equiv
\left[ \begin{array}{cc}
\left( s+1+i \bar{\gamma} \right)  &
(1+i \bar{\gamma})  \\
(1-i \bar{\gamma})   &
\left( s+1-i \bar{\gamma} \right) 
\end{array} \right].
\label{eq:LSCoh_pertmat2}
\end{equation}

\noindent
Equation (\ref{eq:LSCoh_alg2}) implies that either (i) $\mbox{det} \Upsilon = 0$,
or (ii)$\langle \delta \hat{Z} \rangle = \langle \delta \hat{Z}^* \rangle = 0$.
Possibility (i) gives $s(s+2)=0$, yielding
$s=0$ and $s=-2$. Physically, the neutrally stable root, $s=0$, corresponds
to a uniform, rigid rotation of the phases of all the $\hat{Z}_j$.
If possibility (ii) applies, Eqs. (\ref{eq:LSCoh_pert_Z}) 
and (\ref{eq:LSCoh_pert_Zconj}) become

% ----------------------------------------------------------------- LSCoh_alg3
\begin{equation}
\Psi 
\left[ \begin{array}{c}
\delta \hat{Z}_j  \\
\delta \hat{Z}_j^* 
\end{array} \right] 
=0,
\label{eq:LSCoh_alg3}
\end{equation}

\noindent
where
% -------------------------------------------------------------- LSCoh_pertmat3
\begin{equation}
\Psi \equiv
\left[ \begin{array}{cc}
\left( s+2+i \bar{\gamma} \right)  &
(1+i \bar{\gamma})  \\
(1-i \bar{\gamma})   &
\left( s+2-i \bar{\gamma} \right) 
\end{array} \right].
\label{eq:LSCoh_pertmat3}
\end{equation}

\noindent
Since $\mbox{det} \Psi = (s+1)(s+3)$, we obtain two additional roots
$s=-1$ and $s=-3$. Because the allowed perturbations in case (ii) are
restricted to lie in the $2(N-1)$ dimensional space specified by the two constraints,
$\langle \delta \hat{Z} \rangle =0$ and $\langle \delta \hat{Z}^* \rangle = 0$,
the multiplicity of each of the roots $s=-1$ and $s=-3$ is $N-1$. Since there is
no root with $Re(s) >0$, the equilibrium is stable. Hence, the 
single-cluster locked state is stable.

% ============================================================================
\section{Large coupled Landau-Stuart oscillator networks in the strong coupling limit: Cluster states} 
\label{sec:CS_strong_couple}

% ====================================================================== single cluster
\subsection{Regime of global attraction for the single-cluster locked state}\label{sec:regime_singlecluster}
In Sec. \ref{sec:above_instab} we numerically suggest the tendency of system (\ref{eq:LanStuNet3}) 
to form a single-cluster locked
state when $\Gamma$ is sufficiently large, and in Sec. \ref{sec:LS_strong_couple} we have shown
that such an attractor always exists at large $\Gamma$. 
In this section, we give a sufficient condition for this
state to be the {\it only} attractor of the system.
Consider any two oscillators, $m$ and $n$, in (\ref{eq:LanStuNet3}). Let 
% ----------------------------------------------------------------------------- LanNet3_cen_diff
\begin{subequations}
\begin{align}
\bar{\hat{z}} &= \frac{\hat{z}_m + \hat{z}_n}{2}, \label{LanNet3_cen} \\
\delta &= \frac{\hat{z}_m - \hat{z}_n}{2}. \label{LanNet3_diff}
\end{align}
\label{eq:LanNet3_cen_diff}
\end{subequations}

\noindent
Then the dynamical equation for the separation between the two
oscillators, $\delta$, can be immediately derived from Eq. (\ref{eq:LanStuNet3}) as
% ---------------------------------------------------------------------------- LanNet3_diffdyn
\begin{equation}
\frac{d}{d\hat{t}} \delta = -(1+i\bar{\gamma}) \left( 2 |\bar{\hat{z}}|^2 \delta + \delta^* \bar{\hat{z}}^2
                                            + |\delta|^2 \delta \right).
\label{eq:LanNet3_diffdyn}
\end{equation}

\noindent
Letting $\delta = |\delta| e^{i \nu_1}$ and $\bar{\hat{z}} = |\bar{\hat{z}}| e^{i \nu_2}$, substitution
into (\ref{eq:LanNet3_diffdyn}) yields
% -------------------------------------------------------------------------- LanNet3_diffdyn2
\begin{equation}
\frac{d}{d\hat{t}} |\delta| = -|\bar{\hat{z}}|^2 |\delta| [2+\cos(\nu_3) + \bar{\gamma} \sin(\nu_3)]
                                            - |\delta|^3, 
\label{eq:LanNet3_diffdyn2}
\end{equation}

\noindent
where $\nu_3 = 2(\nu_1-\nu_2)$. The trigonometric terms on the right hand side of
Eq. (\ref{eq:LanNet3_diffdyn2}) can be combined, giving
% ---------------------------------------------------------------------- LanNet3_diffdyn3
\begin{equation}
\frac{d}{d\hat{t}} |\delta| = -|\bar{\hat{z}}|^2 |\delta|[2 + \sqrt{1+\bar{\gamma}^2} \cos(\nu_4-\nu_3)]
                                            - |\delta|^3,
\label{eq:LanNet3_diffdyn3}
\end{equation}

\noindent
where $\nu_4$ is determined by the conditions $\cos(\nu_4) = 1 / \sqrt{1+\bar{\gamma}^2}$
and $\sin(\nu_4) = \bar{\gamma} / \sqrt{1+\bar{\gamma}^2}$.
From Eq. (\ref{eq:LanNet3_diffdyn3}), we see that $d |\delta| / d\hat{t} < 0$ if
$2 > \sqrt{1+\bar{\gamma}^2}$, or equivalently
$\bar{\gamma} < \sqrt{3}$.  Thus, if $\bar{\gamma} < \sqrt{3}$, the attractor of
system (\ref{eq:LanStuNet3}) must occur as a single-cluster, and, as shown 
in Sec. \ref{sec:LS_strong_couple}, a single-cluster attractor must be a locked
state (\ref{eq:LSCoh_asympsol}).
However, we shall soon see
that if $\bar{\gamma} > \sqrt{3}$, then Eq. 
(\ref{eq:LanStuNet3}) has the possibility of attracting solutions other than the single-cluster 
locked state. This technique [Eqs. (\ref{eq:LanNet3_cen_diff})-(\ref{eq:LanNet3_diffdyn3})]
can also be employed for related problems such as when the coupling is complex,
$\Gamma \rightarrow \Gamma e^{i \chi}$, or when the coupling is in the form in 
Eq. (\ref{eq:LanStu_oldw}).

% ========================================================== multiple clusters
\subsection{Cluster States}\label{sec:Cluster_states}
We now wish to investigate the possible existence of attracting states for 
(\ref{eq:LanStuNet3}) composed of $C \geq 2$ clusters,
where each cluster is
labeled by a subscript $c=1,2,\cdots, C$. Each cluster $c$ has $N_c$ oscillators in identical
states $Z_c$, such that if oscillators $i$ and $j$ are in cluster $c$, then
$\hat{z_i}=\hat{z_j}=Z_c$, and $N_1 + N_2 + \cdots N_C=N$. Letting $\xi_c = N_c/N$ be the
fraction of oscillators in cluster $c$, we have that 

% -------------------------------------------------------------------- CS_ordp
\begin{equation}
\langle Z \rangle = \xi_1 Z_1 + \cdots + \xi_C Z_C,
\label{eq:CS_ordp}
\end{equation}

\noindent
and that Eq. (\ref{eq:LanStuNet3}) reduces to $C$ equations for the $C$ complex cluster
variables $Z_c$ ($c=1,2, \cdots, C$),

% -------------------------------------------------------------------- CS_LSNet
\begin{equation}
\frac{d Z_c}{d\hat{t}} = -(1+i\bar{\gamma}) |Z_c|^2 Z_c + \sum_{c=1}^C \xi_c Z_c.
\label{eq:CS_LSNet}
\end{equation}

\noindent
Two questions pertaining to such cluster states are (i) what are the attractors of 
(\ref{eq:CS_LSNet}), and (ii) given an attractor of (\ref{eq:CS_LSNet}), are the
clusters internally stable? Question (ii) asks whether, if we consider oscillators
in cluster $c$ and individually independently perturb each of them from
their common value $Z_c$, do they relax back to a common value? (This question 
was considered for the one-cluster locked state in Sec. \ref{sec:LS_strong_couple}.)

The question of the existence of cluster state attractors is a general one applicable
to any large system of identical dynamical units that are coupled via a 
global field (e.g., in our case $\langle Z \rangle$). In particular, this type of
consideration was introduced by Kaneko who considered coupled maps 
\cite{Kaneko1,Kaneko2}.

In our numerical experiments we have always found that, at long time, the solutions
of (\ref{eq:LanStuNet3}) settles into a finite number of clusters $Z_c(t)$. We caution
that this does not necessarily mean that attractors of (\ref{eq:LanStuNet3}) always
occur in clusters, but rather that we have so far not found non-clustered
long-time states. References \cite{Hakim}-\cite{Nakagawa2} consider
a globally coupled Landau-Stuart system with {\it homogeneous} parameters
across all oscillators (our (\ref{eq:LanStuNet3}) is a special case) and find both
clustered state attractors and ``scattered state'' attractors, where
by ``scattered states'' they mean that, at any given time, no two oscillators have exactly the same
state. We, however, have not seen scattered state attractors, and we conjecture that
they do not exist for (\ref{eq:LanStuNet3}). Along these lines, we now show a partial
result implying that scattered states cannot have scattering that is {\it over an area}
in the $z$-plane. That is, in the limit $N \rightarrow \infty$ the distribution function 
$f$ appearing in (\ref{eq:f_evol}) must be singular in the sense that it is concentrated
on a set of zero Lebesgue measure in $z$-space, equivalently ($\rho,\theta$) space. Examples
of zero Lebesgue measure sets are a set of distinct points (like our clusters), a curve,
or a fractal set of dimension between one and two. Indeed the scattered states seen
in the figures of the previous references \cite{Hakim}-\cite{Nakagawa2} 
(e.g., Fig. 1 of Ref. \cite{Nakagawa2}) appear to our eye 
to be either
fractal distributions with dimension near one or distributed along a convoluted curve
(based on Ref. \cite{Yu}, we suspect that the first of these alternatives applies). The demonstration
that $f$ must be singular follows simply from (\ref{eq:f_evol}) by introducing 
$F = f / \rho$ and rewriting (\ref{eq:f_evol}) in the form

% --------------------------------------------------------- f_evol_v2
\begin{equation}
\frac{d}{dt} F = \frac{\partial}{\partial t} F+  \dot{\rho} \frac{\partial}{\partial \rho} F
  + \dot{\theta} \frac{\partial}{\partial \theta} F = 4 \rho^2 F,
\label{eq:f_evol_v2}
\end{equation}

\noindent
where $\dot{\rho}$ and $\dot{\theta}$ are the two quantities in (\ref{eq:f_evol}) appearing
within the square brackets with the linear oscillator parameters, $\alpha$
and $\omega$, set to zero [to correspond to (\ref{eq:LanStuNet3})]. Note too that
$F = F(\rho,\theta,t)$; in particular, unlike our more general set-up,
Eq. (\ref{eq:f_evol}), $F$ does not incorporate distribution in parameters as we have fixed 
$\alpha$ and $\omega$ at zero. 
According to (\ref{eq:f_evol_v2}), following the characteristics,
$d\rho / dt=\dot{\rho}$, $d\theta /dt = \dot{\theta}$, of the partial differential
equation (\ref{eq:f_evol}) with $\alpha = \omega =0$,
$F$ increases continually at the exponential rate 
$4 \rho^2$. Thus, since $\int f d\rho d\theta = \int F \rho d\rho d\theta = 1$,
we immediately conclude that 
in the long time
limit, $F$ and hence $f$ must concentrate on a set of zero area (zero Lebesgue measure) in ($\rho,\theta$) space.
Proof that our system (\ref{eq:LanStuNet3}) does or does not always yield cluster state attractors remains
an open problem.

% =========================================================== C=2
\subsection{Two-cluster states}\label{sec:Two_CS}
For $C=2$, Eq. (\ref{eq:CS_LSNet}) yields

% ---------------------------------------------------------- LanStuNet_CS
\begin{subequations}
\begin{align}
\frac{dZ_1}{d\hat{t}} &= -(1+i\bar{\gamma}) |Z_1|^2 Z_1 
+ \xi_1 Z_1 + \xi_2 Z_2, \label{eq:LanStuNet_CS1} \\
\frac{d Z_2}{d\hat{t}} &= -(1+i\bar{\gamma}) |Z_2|^2 Z_2 
+ \xi_1 Z_1 + \xi_2 Z_2. \label{eq:LanStuNet_CS2}
\end{align}
\label{eq:LanStuNet_CS}
\end{subequations}

\noindent
Letting $Z_1 = \tilde{\rho}_1 e^{i \tilde{\theta}_1}$
and $Z_2 = \tilde{\rho}_2 e^{i \tilde{\theta}_2}$, and defining the relative
phase difference $\tilde{\phi} = \tilde{\theta}_1 - \tilde{\theta}_2$, Eq.
(\ref{eq:LanStuNet_CS}) yields three real equations,

% ---------------------------------------------------------- LanStuNet_CS2
\begin{subequations}
\begin{align}
\frac{d \tilde{\rho}_1}{d\hat{t}} &= \tilde{\rho}_1(\xi_1 - \tilde{\rho}_1^2) + 
                            \xi_2 \tilde{\rho}_2 \cos(\tilde{\phi}), 
                             \label{eq:Lan_CS2_rho1} \\
\frac{d \tilde{\rho}_2}{d\hat{t}} &= \tilde{\rho}_2 (\xi_2 - \tilde{\rho}_2^2) + 
                            \xi_1 \tilde{\rho}_1 \cos(\tilde{\phi}), 
                             \label{eq:Lan_CS2_rho2} \\
\frac{d \tilde{\phi}}{d \hat{t}} &= -\bar{\gamma} (\tilde{\rho}_1^2 - \tilde{\rho}_2^2) - \sin(\tilde{\phi}) 
                            \left[ \xi_2 \frac{\tilde{\rho}_2}{\tilde{\rho}_1} 
                               + \xi_1 \frac{\tilde{\rho}_1}{\tilde{\rho}_2} \right]
                            \label{eq:Lan_CS2_phi}.
\end{align}
\label{eq:LanStuNet_CS2}
\end{subequations}

\noindent
Note that $\tilde{\rho}_1 = \tilde{\rho}_2 = 1$ and $\tilde{\phi}=0$
is a solution of these equations and corresponds to the single-cluster locked state. 
We want to obtain two-cluster
solutions ($Z_1 \neq  Z_2$). 
Although we cannot rule out chaotic or two-frequency quasiperiodic two-cluster
solutions of Eq. (\ref{eq:LanStuNet_CS2}), so far our numerical investigations of 
Eqs. (\ref{eq:LanStuNet_CS2}) have only found fixed point attractors (i.e., 
uniformly rotating two-cluster locked states)
and periodic attractors. In the next two
subsections we discuss the fixed point solutions (Sec. \ref{sec:Two_CS_fixed}) and the 
periodic solutions (Sec. \ref{sec:Two_CS_period}).

% ============================================================ C=2, fixed point
\subsubsection{Two-cluster locked states (fixed point solutions of Eq. (\ref{eq:LanStuNet_CS2}))}
\label{sec:Two_CS_fixed}
When the time derivatives in Eq. (\ref{eq:LanStuNet_CS2}) are all zero, the solutions 
give locked two-cluster fixed point solutions. By setting 
$d/dt = 0$, $x = \tilde{\rho}_1^2$
and $y = \tilde{\rho}_2^2$, elimination of $\cos(\tilde{\phi})$ between Eqs. (\ref{eq:Lan_CS2_rho1})
and (\ref{eq:Lan_CS2_rho2}) gives

% ------------------------------------------------------------- 2CS_hyp
\begin{equation}
\xi_1 \left( x-\frac{\xi_1}{2} \right)^2 - \xi_2 \left( y - \frac{\xi_2}{2} \right)
= \frac{1}{4} ( \xi_1^3 - \xi_2^3 ).
\label{eq:2CS_hyp}
\end{equation}

\noindent
On the other hand, addition of Eqs. (\ref{eq:Lan_CS2_rho1}) and (\ref{eq:Lan_CS2_rho2}), and 
subsequent elimination of the trigonometric factor with that in Eq. (\ref{eq:Lan_CS2_phi}) by the
identity $\sin^2(\tilde{\phi}) + \cos^2(\tilde{\phi}) = 1$ gives

% -------------------------------------------------------------- 2CS_oth
\begin{equation}
(x+y-1)^2 + \bar{\gamma}^2 (y-x)^2 = 
    \left( \xi_2 \sqrt{\frac{y}{x}} + \xi_1 \sqrt{\frac{x}{y}} \right)^2.
\label{eq:2CS_oth}
\end{equation}

\begin{figure}
   \subfigure[]
   {\includegraphics[width=8cm]
      {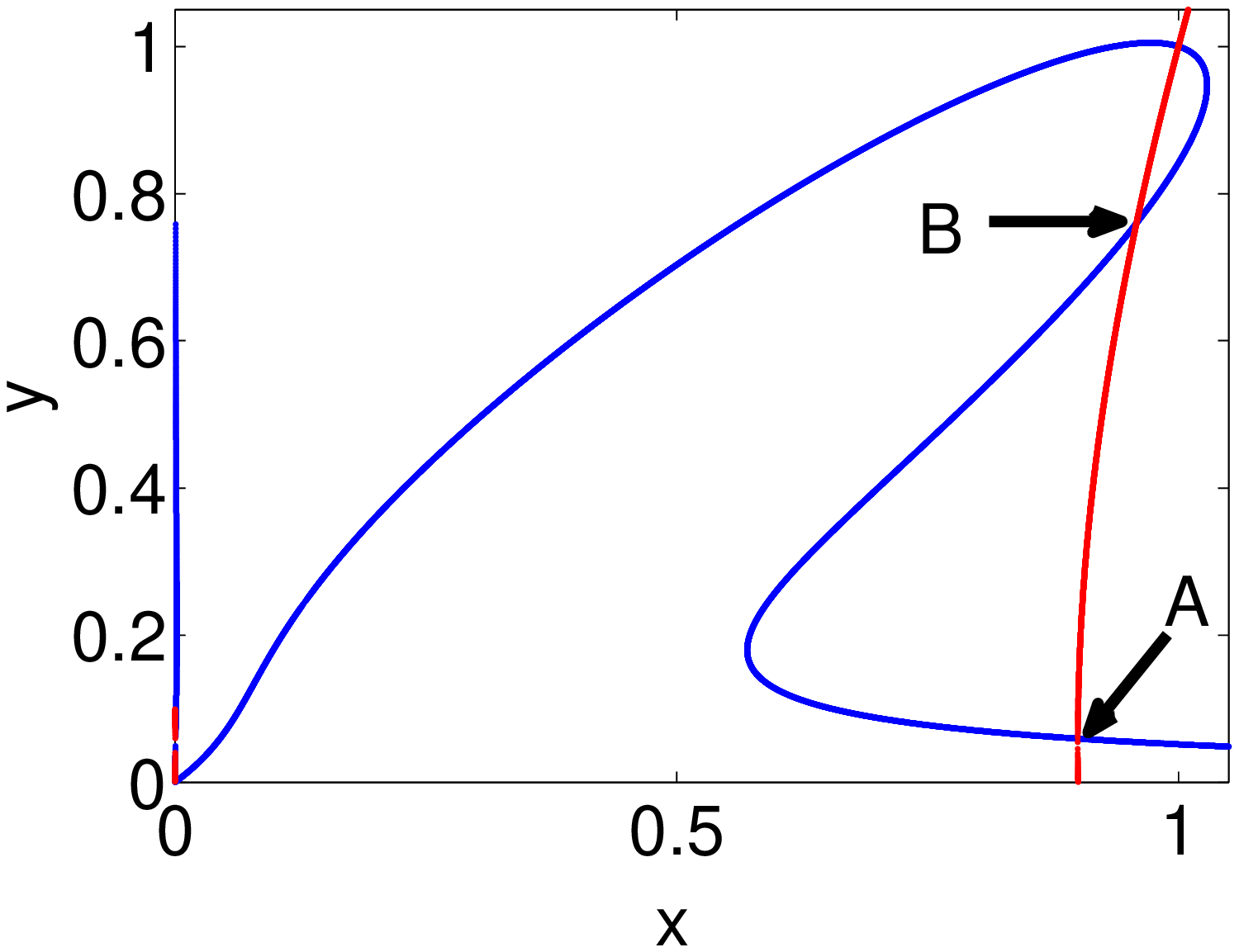}}
   \subfigure[]
   {\includegraphics[width=8cm]
      {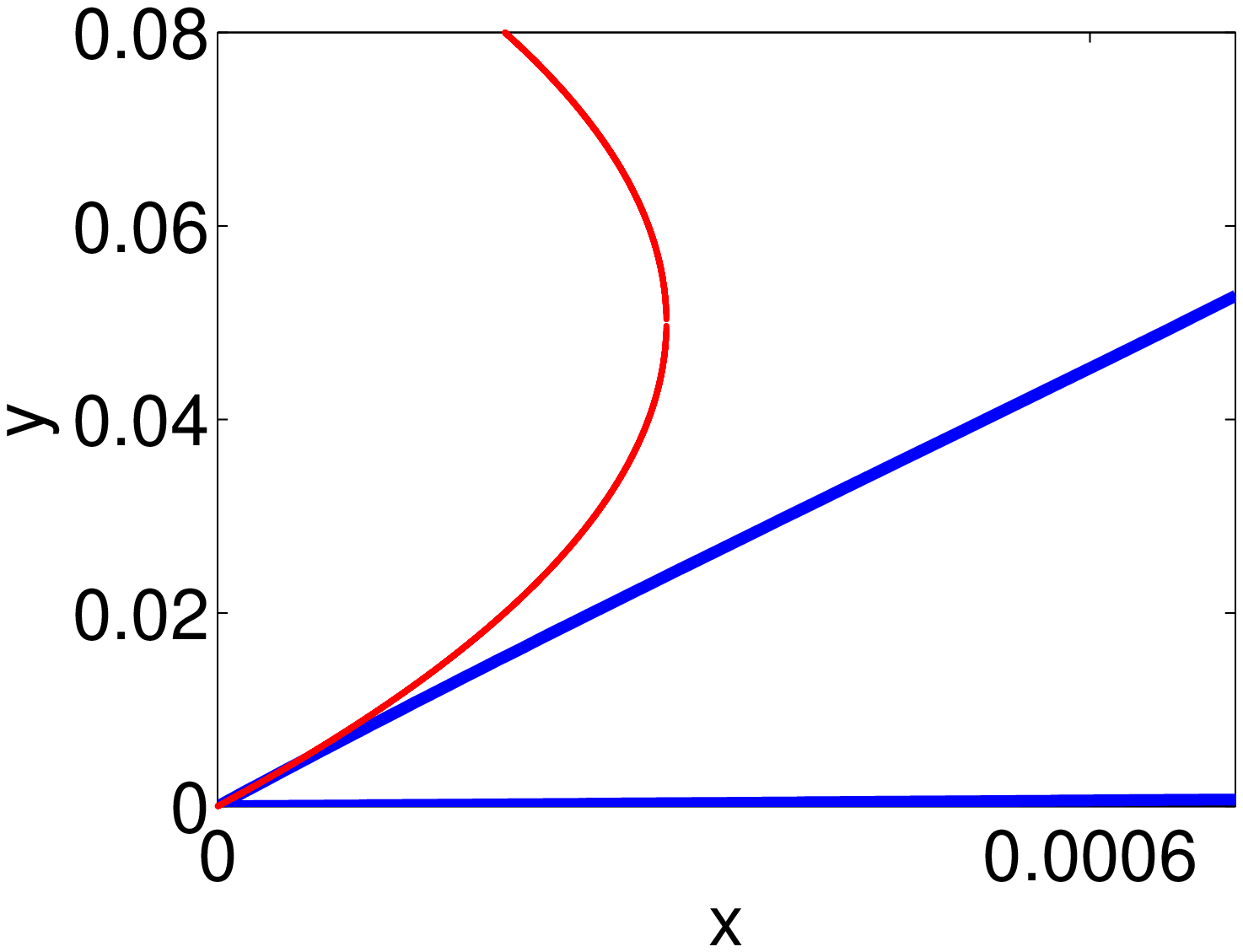}}   
   \caption{(Color online) (a) Two-cluster locked state solutions. The blue line shows the curve for
            Eq. (\ref{eq:2CS_oth}), and the red line shows the curve for Eq. (\ref{eq:2CS_hyp}); 
            parameters: $\xi_1=0.9$, $\bar{\gamma}=4.2$. (b) Blowup of the figure in (a) 
            around $(0,0)$.}
\label{fig:two_cluster_lockedsol}
\end{figure}

\noindent
Two-cluster fixed point solutions are given by the intersecting points of Eqs. (\ref{eq:2CS_hyp})
and (\ref{eq:2CS_oth}). An example shown in Fig. \ref{fig:two_cluster_lockedsol}
corresponding to the parameters $\xi_1 = 0.9$ and $\bar{\gamma}=4.2$. There are altogether four
intersecting points, but two of them, namely $(0,0)$ and $(1,1)$, do not correspond to the
answers we seek ($(0,0)$ is the unstable incoherent state and $(1,1)$ is the one-cluster locked
state solution). For the other two solutions, indicated as $A$ and $B$ in 
Fig. \ref{fig:two_cluster_lockedsol}, we find that
$B$ corresponds to an unstable fixed point, while $A$ is stable, and our numerical tests on
Eqs. (\ref{eq:LanStuNet_CS2}) and (\ref{eq:LanStuNet3}) show that $A$ satisfies both types of
stability, (i) and (ii) mentioned at the end of Sec. \ref{sec:Cluster_states}. Thus $A$ is an
attractor.

% fffffffffffffffffffffffffffffffffffffffffffffffffffffffffffffffffffffffffffffffffffffffffff

% ============================================================ C = 2, periodic
\subsubsection{Two-cluster periodic solutions of Eq. (\ref{eq:LanStuNet_CS2})}\label{sec:Two_CS_period}
In general, the two-cluster periodic solutions of Eq. (\ref{eq:LanStuNet_CS2}) are hard to 
obtain analytically. 
However, in the case of large $\bar{\gamma}$, we can proceed using 
perturbation theory. If
$\bar{\gamma} \gg 1$ and $\tilde{\rho}_1^2 - \tilde{\rho}_2^2$ is not small, 
then $d\tilde{\phi} / d\hat{t}$ is
very large, so $\cos(\tilde{\phi})$ is rapidly varying. Thus, to lowest order in $\bar{\gamma}^{-1}$,
we can neglect the $\cos(\tilde{\phi})$ terms in the equations for $d\tilde{\rho}_1/d\hat{t}$
and $d\tilde{\rho}_2/d\hat{t}$,

% ---------------------------------------------------------- Lan_CS_apprho
\begin{equation}
\frac{d\tilde{\rho}^{(0)}_{1,2}}{d\hat{t}} = (\xi_{1,2} - \tilde{\rho}^{(0)^2}_{1,2}) \tilde{\rho}^{(0)}_{1,2},
\label{eq:Lan_CS_apprho}
\end{equation}

\noindent
and $\tilde{\rho}^{(0)}_1$ and $\tilde{\rho}^{(0)}_2$, are attracted to $\tilde{\rho}^{(0)}_1 = \sqrt{\xi_1}$,
$\tilde{\rho}^{(0)}_2 = \sqrt{\xi_2}$, respectively. Thus we have

% -------------------------------------------------------- Lan_CS_appphi
\begin{equation}
\frac{d\tilde{\phi}}{d\hat{t}} \approx -\bar{\gamma} (\xi_1-\xi_2) - 
                       \frac{\xi_1^2+\xi_2^2}{\sqrt{\xi_1 \xi_2}} \sin(\tilde{\phi}),
\label{eq:Lan_CS_appphi}
\end{equation}

\noindent
which, for large $\bar{\gamma}$ has the solution,
% -------------------------------------------------------- Lan_CS_appphisol
\begin{equation}
\tilde{\phi} \approx \tilde{\phi}_0 - \bar{\gamma} (\xi_1-\xi_2)\hat{t} + 
             \frac{\xi_1^2+\xi_2^2}{\sqrt{\xi_1 \xi_2}} \cos[\tilde{\phi}_0-\bar{\gamma}(\xi_1-\xi_2)\hat{t}].
\label{eq:Lan_CS_appphisol}
\end{equation}

\noindent
To next order in $\bar{\gamma}^{-1}$, we write
$\tilde{\rho}_{1,2}=\tilde{\rho}^{(0)}_{1,2}+\tilde{\rho}^{(1)}_{1,2}$
where the perturbation $\tilde{\rho}^{(1)}_1$ to the lowest order quantity $\tilde{\rho}^{(0)}_1$ satisfies

% --------------------------------------------------------- Lan_CS_apprho1a
\begin{equation}
\frac{d\tilde{\rho}^{(1)}}{d\hat{t}} = -2 \xi_1 \tilde{\rho}^{(1)}_1 
   + \xi_2^{3/2}\cos[\tilde{\phi}_0 - \bar{\gamma}(\xi_1-\xi_2)\hat{t}].
\label{eq:Lan_CS_apprho1a}
\end{equation}

\noindent
The homogeneous solution of the above equation decays as $e^{-2\xi_1 \hat{t}}$,
and thus does not contribute to the time periodic attractor. For large
$\bar{\gamma}$ the inhomogeneous term varies with a period $2\pi / [\bar{\gamma}(\xi_1-\xi_2)]$
which, since $\bar{\gamma}$ is large, is much shorter than the damping time $(2\xi_1)^{-1}$.
Thus for calculating the inhomogeneous solution, we may neglect
the term $2 \xi_1 \tilde{\rho}^{(1)}_1$. This yields

% ----------------------------------------------------- Lan_CS_rho1a
\begin{equation}
\tilde{\rho}^{(1)}_1 (\hat{t}) = -\frac{\xi_2^{3/2}}{\bar{\gamma}(\xi_1-\xi_2)} 
\sin[\tilde{\phi}_0-\bar{\gamma}(\xi_1-\xi_2)\hat{t}],
\label{eq:Lan_CS_rho1a}
\end{equation}

\noindent
and similarly

% ----------------------------------------------------- Lan_CS_rho1b
\begin{equation}
\tilde{\rho}^{(1)}_2 (\hat{t}) = -\frac{\xi_1^{3/2}}{\bar{\gamma}(\xi_1-\xi_2)} 
\sin[\tilde{\phi}_0-\bar{\gamma}(\xi_1-\xi_2)\hat{t}].
\label{eq:Lan_CS_rho1b}
\end{equation}

\begin{figure}
   {\includegraphics[width=8cm]
      {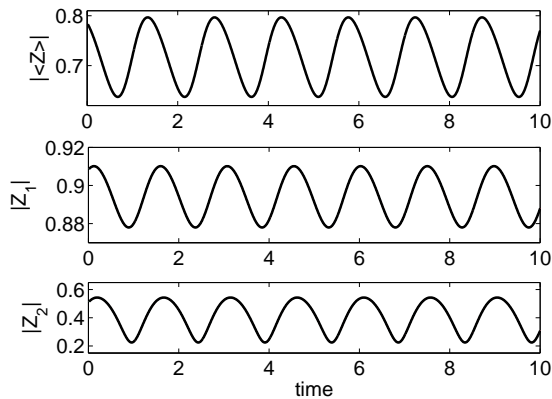}}
   \caption{Simulation study of a two-cluster periodic state; 
            parameters: $N=50000$, $\xi_1=0.8$, $\bar{\gamma}=7.2$.}
\label{fig:two_cluster_periodsol}
\end{figure}

\noindent
Thus $\rho^{(1)}_{1,2}$ are indeed small compared to $\rho^{(0)}_{1,2}$, if
$\xi_1 \neq \xi_2$ and $\bar{\gamma}$ is sufficiently large. Hence we obtain a two
cluster time periodic state whose frequency is, to lowest order, $\bar{\gamma} (\xi_1-\xi_2)$.
Figure \ref{fig:two_cluster_periodsol} shows long-time results of a simulation of
Eq. (\ref{eq:LanStuNet3}) with
$50000$ oscillators, and with parameters $\xi_1=0.8$
and $\bar{\gamma}=7.2$, for a periodic attractor. Comparison of these results with the 
approximate analytical solution above shows good agreement.

% ffffffffffffffffffffffffffffffffffffffffffffffff

In order to see why this solution represents an attractor of the full $N$ dimensional
system (\ref{eq:LanStuNet3}), we consider its Lyapunov exponents.
To lowest order the individual clusters are uncoupled locked states.
We have already shown (Sec. \ref{sec:LS_strong_couple}) that a single-cluster locked
state of a system of $N$ oscillators has $N-1$ negative exponents
(having possible values $-1,-2,-3$) and one zero exponent that corresponds
to a rigid rotation in the complex plane of the entire system of 
$N$ oscillators (i.e., $\hat{z}_j \rightarrow \hat{z}_j \exp(i \Phi)$ for constant $\Phi$).
Thus, to lowest order in our $\bar{\gamma}^{-1}$ expansion, there are 
$(N_1-1)+(N_2-1)=N-2$ negative Lyapunov exponents and two zero Lyapunov exponents.
Now introducing small coupling between the clusters (i.e., finite $\bar{\gamma}$), 
the negative exponents will
be slightly perturbed by an amount $\bar{\gamma}^{-1} \ll 1$ and hence will remain
negative. The only danger of instability is that one of the two zero exponents might be
perturbed to be a positive number of order $\bar{\gamma}^{-1}$. However, this cannot be
the case, because the full system must have two zero exponents, and thus the two
zero exponents of the lowest order uncoupled approximation are preserved. To see this,
we note that there is one zero exponent corresponding to a rigid rotation of the entire
system of $N = N_1 + N_2$ oscillators. Note that this zero exponent is 
not present in the three ODE's, Eq. (\ref{eq:LanStuNet_CS2}), for 
$\tilde{\rho}_1, \tilde{\rho}_2$ and $\tilde{\phi}$, since a rigid rotation
$(\tilde{\theta}_1 \rightarrow \tilde{\theta}_1 + \Phi, \tilde{\theta}_2 
\rightarrow \tilde{\theta}_2 + \Phi)$ does not
change $\tilde{\phi} = \tilde{\theta}_1-\tilde{\theta}_2$. Another zero exponent results from the
fact that the time periodic flow, Eq. (\ref{eq:LanStuNet_CS2}), has a zero exponent corresponding
to displacement along its orbit. Thus we conclude that our large-$\bar{\gamma}$,
two-cluster, states are attractors.

% ================================================ CS_Cgeq3
\subsection{Cluster-states with $C \geq 3$} \label{sec:CS_Cgeq3}

\subsubsection{Do locked state attractors with three or more clusters exist?}\label{sec:CS_3clu_locked}
The above implies that, at large $\Gamma$, we can have both
two-cluster and single-cluster locked state attractors. A natural question is whether
large-$\Gamma$ locked state attractors with $C > 2$ clusters are possible. We now
give a partial answer to this question by ruling out the 
possibility of locked states composed of more than three clusters. 
To rule
out $C > 3$ locked state solutions of Eq. (\ref{eq:LanStuNet3}), we substitute
the locked-state ansatz $Z_c(\hat{t}) = \tilde{\rho}_c e^{i \tilde{\theta}_{c0}} e^{-i \Omega \hat{t}}$
into (\ref{eq:LanStuNet3}), where $\tilde{\rho_c}>0$ and $\tilde{\theta}_{c0}$ are time independent
real constants. This yields 

% -------------------------------------------------------------- CS_fixedpt_sub
\begin{subequations}
\begin{align}
\langle Z(\hat{t}) \rangle = \langle Z(0) \rangle e^{-i \Omega \hat{t}}, 
\hspace{2mm} \mbox{and} \label{eq:CS_fixedpt_sub1} \\
[(1+i\bar{\gamma})\tilde{\rho}_c^2 - i \Omega] \tilde{\rho}_c e^{i \theta_{c0}} = \langle Z(0) \rangle
\label{eq:CS_fixedpt_sub2}
\end{align}
\label{eq:CS_fixedpt_sub}
\end{subequations}

\noindent
Multiplying Eq. (\ref{eq:CS_fixedpt_sub2}) by its complex conjugate, we obtain 

% --------------------------------------------------------------- CS_fixedpt_poly
\begin{equation}
\tilde{\rho}_c^2 [\tilde{\rho}_c^4 +
(\bar{\gamma} \tilde{\rho}_c^2 - \Omega)^2] = |\langle Z(0) \rangle|^2.
\label{eq:CS_fixedpt_poly}
\end{equation}

\noindent 
A particular state
corresponds to particular values of $\Omega$ and $|\langle Z(0) \rangle|$. Thus
Eq. (\ref{eq:CS_fixedpt_poly}) must be satisfied for each individual cluster
$c$ composing the state for the same values of $\Omega$ and $|\langle Z(0) \rangle|$.
Since (\ref{eq:CS_fixedpt_poly}) is a cubic equation for $\tilde{\rho}_c^2$, there can
be at most three real values of $\tilde{\rho}_c > 0$. Furthermore, Eq. (\ref{eq:CS_fixedpt_sub})
uniquely determines the value of $\tilde{\theta}_{c0}$ for each value of $\tilde{\rho}_c$.
We, therefore, conclude that large-$\Gamma$, locked, cluster states with 
$C > 3$ cannot occur. This leaves open the question of whether or not three cluster 
locked state attractors exist. In this regard, we note that in our, admittedly limited, series
of numerical experiments we have so far not seen such attractors.

\subsubsection{Periodic, quasiperiodic and chaotic attractors for $C \geq 3$}\label{sec:CS_3clu_other}
Considering the $C$ complex ODE's for the $C$ cluster states, Eq. (\ref{eq:CS_LSNet}), 
and again
introducing a polar representation, $Z_c = \tilde{\rho}_c \exp(i \tilde{\theta}_c)$,
we obtain a $2C-1$ dimensional dynamical system of $C$ 
real equations for $d \tilde{\rho}_c / d\hat{t}$
and $C-1$ real equations for 
$d \tilde{\phi}_c / d\hat{t}$ where $\tilde{\phi}_c = \tilde{\theta}_c - \tilde{\theta}_1$
and $c=2,\cdots C$. Again taking $\bar{\gamma} \gg 1$, we find $C-1$ lowest order
angle evolutions,

% ------------------------------------------------------------------------ CS_phi
\begin{equation}
\frac{d \tilde{\phi}_c^{(0)}}{d \hat{t}} = \Delta \omega_c, \hspace{2mm} \Delta \omega_c = -\bar{\gamma}
(\xi_c - \xi_1).
\label{eq:CS_phi}
\end{equation}

\noindent
Assuming that the set of $C-1$ frequencies $\{ \Delta \omega_c \}$ are irrationally related in the 
sense that the equation,
% ----------------------------------------------------------------------- IrrationR
\begin{equation}
\sum_{c=1}^{C-1} m_c \Delta \omega_c= 0,
\label{eq:IrrationR}
 \end{equation}
 
\noindent
has no solution where the $m_c$ are positive or negative integers except for the trivial solution
where $m_c = 0$ for all $c$, then we can think of the lowest order solution as being 
$(C-1)$-quasiperiodic in the $2C-1$ cluster-state phase space $\{ \rho_1, \cdots, \rho_C;
\phi_2, \cdots, \phi_C \}$. Application of perturbation theory in the small parameter $\bar{\gamma}^{-1}$
is mathematically equivalent to the problem of investigating the introduction of small coupling 
between $C-1$ oscillators.

For example, for three clusters, we have the possibility of two-frequency quasiperiodic motion,
and the possibly analogous problem of introducing small {\it generic}
coupling between two periodic oscillators 
was originally addressed by Arnold \cite{Arnold} in his study of the circle map, 
$\Xi_{n+1} = (2\pi R_0 + \Xi_n + \kappa \sin \Xi_n) \mod 2\pi$, where $R_0$ denotes the
rotation number for $\kappa = 0$ ($R_0$ is analogous to $\Delta \omega_1 / \Delta \omega_2$
in our case, and $\kappa$ is analogous to $\bar{\gamma}^{-1}$). Arnold's work resolved the
problem of the convergence of perturbation theory of coupled oscillators
which is plagued by the proliferation of small denominators in higher and higher
order terms in the perturbation series. Results, both analytical (as by Arnold \cite{Arnold})
and numerical, show that for $\kappa < 1$, attracting quasiperiodic motion continues
to exist on a positive Lebesgue measure of the parameter space (in our case, the 
parameter-space is $\{\xi_c;\bar{\gamma}\}$), but is structurally unstable: for any parameter
set yielding two-frequency quasiperiodicity, one can find an arbitrarily close-by
set of parameters where the motion is periodic. Alternatively, one can say that 
two-frequency quasiperiodic attractors exist on a positive Lebesgue measure
Cantor set in parameter-space, while periodic attractors exist on the complement of this set
which is an open set
(e.g., see \cite{Ott} for further discussion).

% ffffffffffffffffffffffffffffffffffffffffffffff
\begin{figure}
   \subfigure[]
   {\includegraphics[width=8cm]
      {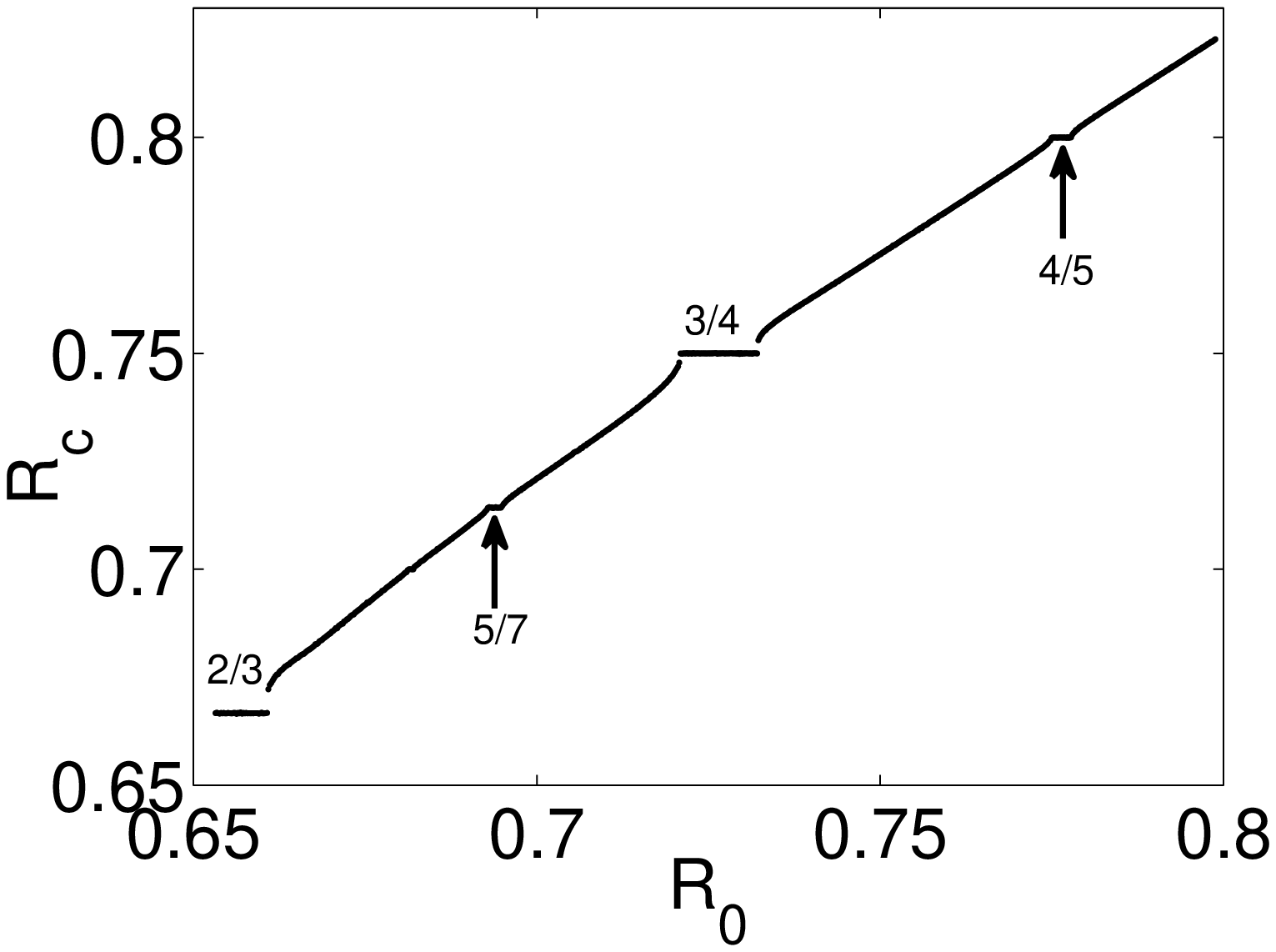} \label{fig:3CS_staircase_whole}}
   \subfigure[]
   {\includegraphics[width=8cm]
      {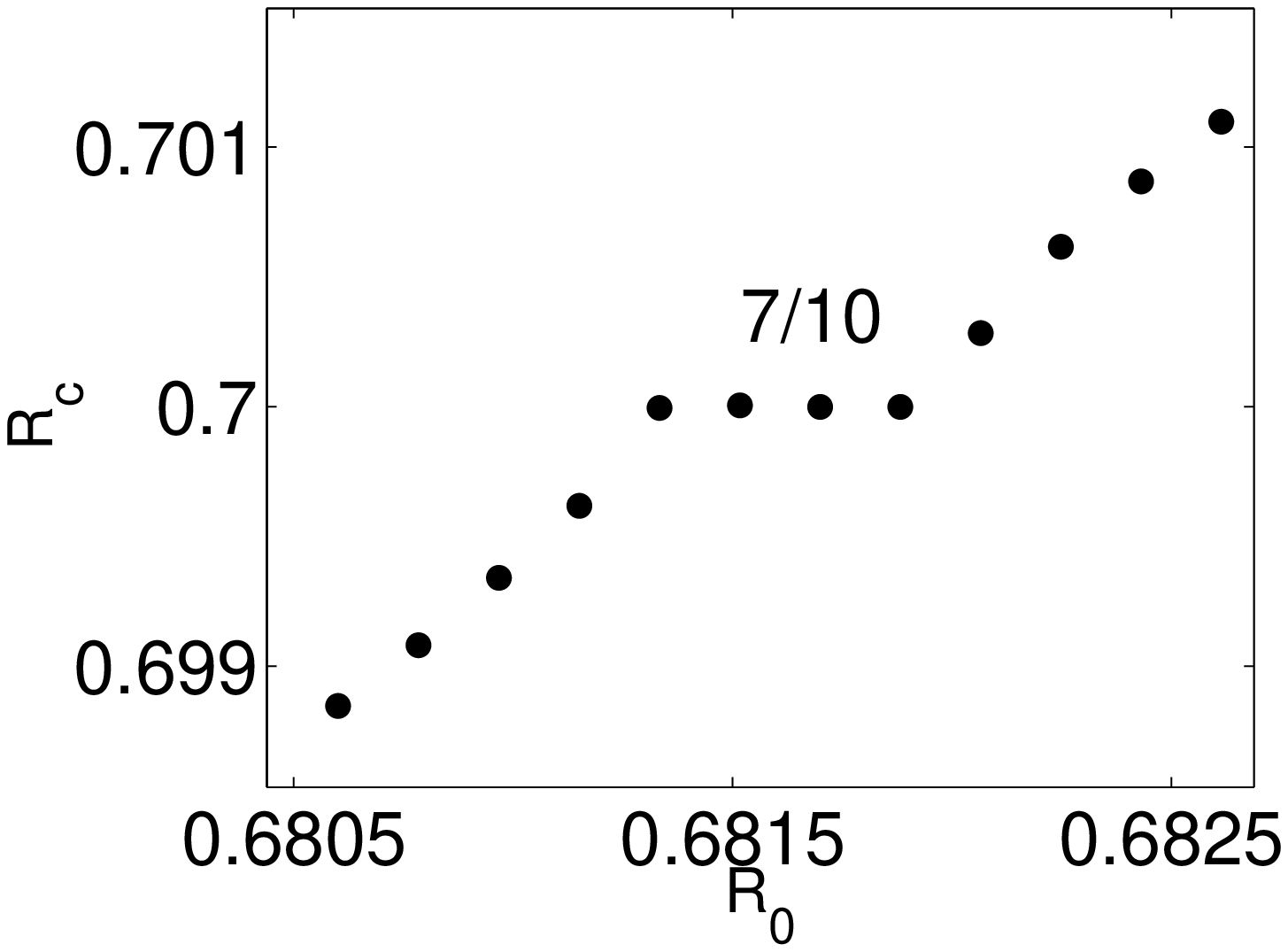} \label{fig:3CS_staircase_blowup1}}
   \subfigure[]
   {\includegraphics[width=8cm]
      {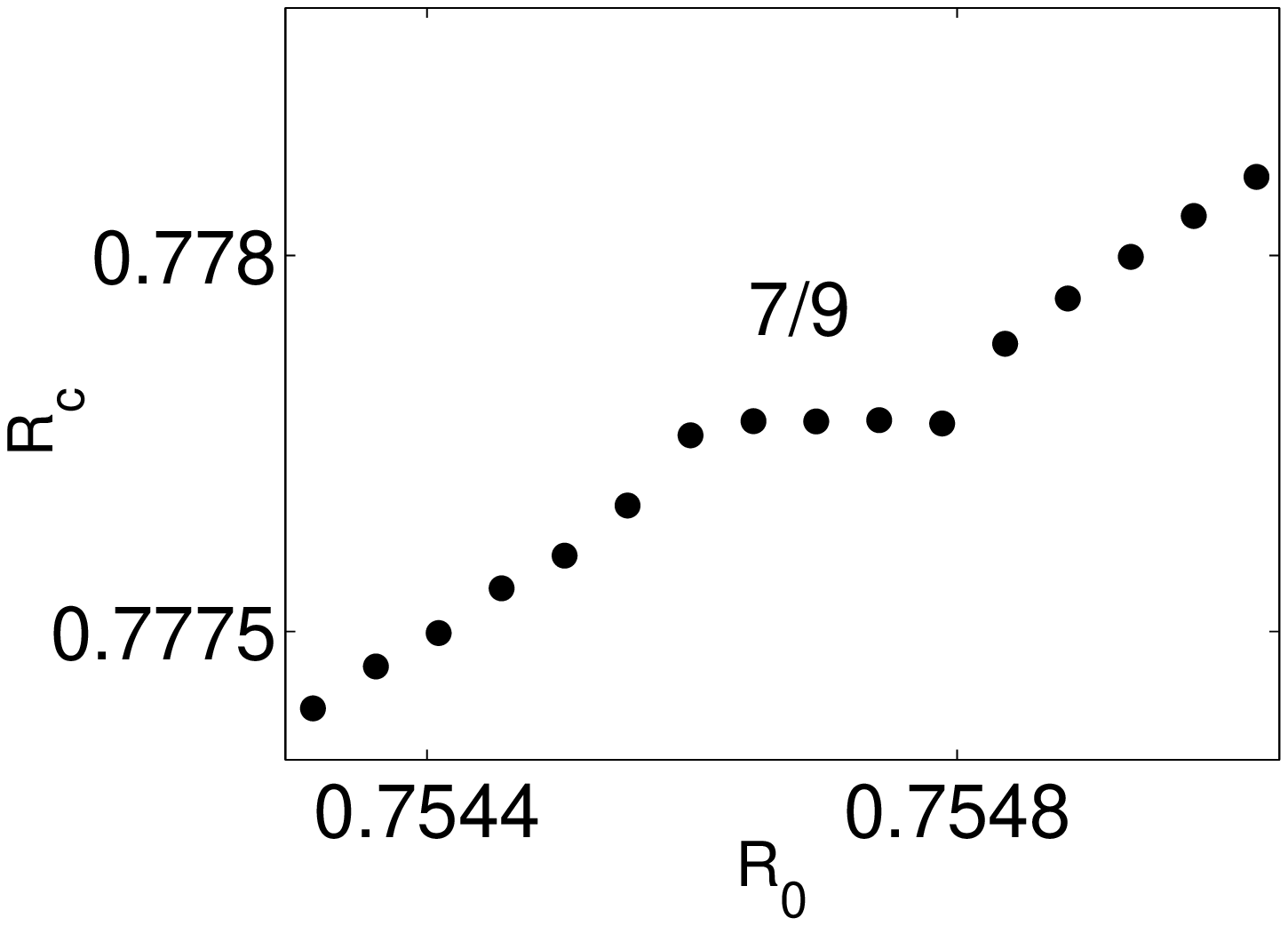} \label{fig:3CS_staircase_blowup2}}

   \caption{(a) Rotation number $R_c$ from numerical solutions versus the infinite $\bar{\gamma}$
            rotation number $R_0$ for $C=3, \bar{\gamma}=30, \xi_3=0.2$, and $\xi_1$ varied from
            $0.497$ to $0.533$ with $\xi_2 = 1-\xi_1-\xi_3$. (b), (c) Blowup of two regions of 
            the figure in (a).}
\label{fig:3CS_staircase}
\end{figure}

Figure \ref{fig:3CS_staircase} shows evidence supporting this scenario. The figure shows
the result of numerical computations of (\ref{eq:CS_LSNet})
for the rotation number defined by
% ----------------------------------------------------------- 3CS_RotationNo
\begin{equation}
R_c = \lim_{T \rightarrow \infty} \frac{\theta_2(T)-\theta_1(T)}{\theta_3(T)-\theta_1(T)},
\label{eq:3CS_RotationNo}
\end{equation}

\noindent
versus the rotation number at infinite $\bar{\gamma}$,

% ------------------------------------------------------------ 3CS_RotationNo_inf
\begin{equation}
R_0 = \frac{\xi_2-\xi_1}{\xi_3-\xi_1},
\label{eq:3CS_RotationNo_inf}
\end{equation}

\noindent
for $C=3, \bar{\gamma}=30, \xi_3=0.2, \xi_1$ varied from $0.497$ to $0.533$,
and $\xi_2 = 1-\xi_1-\xi_3$. A classic ``devil's staircase'' pattern is clearly
observed, with periodic orbits corresponding to frequency-locking plateaus at 
rational rotation numbers of $R_c = 2/3, 5/7, 3/4$, and $4/5$. The blowups,
shown in Figs. \ref{fig:3CS_staircase_blowup1} and \ref{fig:3CS_staircase_blowup2}, 
reveal further plateaus at $R_c = 7/10$
and $7/9$, suggesting that (as in Arnold's circle map) plateaus exist for all rational
numbers $p/q$ ($p, q$ incommensurate integers) with the plateau widths decreasing
as $q$ increases.

The case $C-1 \geq 3$ adds a new ingredient. Again, as $\bar{\gamma}^{-1}$ increases
from zero ($\bar{\gamma}$ becomes finite),
($C-1$)-quasiperiodic attractors generally 
persist on a positive Lebesgue measure Cantor set in parameter space, but
the open complement of this Cantor set now typically contains a variety
of other types of attractors, including periodic attractors (as for $C=3$), chaotic
attractors, and $M$-frequency quasiperiodic attractors with $M < C-1$ 
(\cite{Newhouse}-\cite{Grebogi2}). In particular, the generic existence
of structurally stable chaotic attractors resulting from perturbations 
to $P$-frequency quasiperiodic attractors
for $P \geq 3$ has been proven by Newhouse, Ruelle and Takens \cite{Newhouse,Ruelle}.

Note that all the motions, including the chaotic motions, referred to above 
occur in the context of Eq. (\ref{eq:CS_LSNet}) and are therefore motions of the
$C$ clusters. Also note that in our discussion above of solutions of Eq. (\ref{eq:CS_LSNet}),
we have treated the parameters $\{ \xi_c \}$ as continuous. While this is formally
allowed for Eq. (\ref{eq:CS_LSNet}), we emphasize that if we consider
that (\ref{eq:CS_LSNet}) is derived from system (\ref{eq:LanStuNet3}) with finite $N$,
then $\xi_c = N_c / N$ can only take on discrete values (although they may be very dense 
for large $N$).

% ======================================================================== conclusions
\section{Discussion and Conclusions} \label{sec:conclusions}
In this paper we have studied some properties of large 
all-to-all coupled Landau-Stuart
oscillator networks. 
The motivation for studying this class of systems is to reveal possible
generic behaviors of large systems of coupled oscillators where the oscillators
have both amplitude and phase degrees of freedom. 

In the first half of this paper (Secs. \ref{sec:linsta}-\ref{sec:freqdist_effect}),
motivated by experiments reported in Ref. \cite{Taylor}, 
we studied stability of the incoherent state ($\langle z \rangle = 0$) 
and determined the stability / instability boundary for different
cases. First, we studied
networks with spread in the distributions of the natural oscillator frequencies 
$\omega$ and the 
linear amplitude growth parameter $\alpha$, but with no nonlinear
frequency shift contribution, i.e., $\gamma_i=0$ for all oscillators $i$. 
Second, we studied
networks with no spread in the distribution $\alpha$, but with
a constant nonlinear frequency shift parameter $\bar{\gamma}$ for all oscillators. 
After establishing a mathematical framework to determine the stability / instability boundary, 
we characterized the changes in 
the stability / instability boundary that these modifications cause. 
First, we found that a spread $\delta \alpha$ in the distribution of $\alpha$ smooths out
the discontinuity at $\alpha = 0$
in the slope of the stability / instability boundary. Second, spread in 
$\alpha$ causes
the minimum of $\Gamma_c$ to shift away from $\bar{\alpha}=0$ to 
$\bar{\alpha} > 0$ when $\delta \alpha > 0$.  Third, increase of
the nonlinear frequency shift parameter $\bar{\gamma}$ monotonically lowers  
$\Gamma_c$.
 
Similar to large networks of phase oscillators
of the Kuramoto type, large networks of Landau-Stuart oscillators 
with small nonlinear frequency shifts
have a tendency to always synchronize into a locked state
exhibiting steady, constant-amplitude sinusoidal motion when the coupling strength is large
enough. 
In order to better understand this behavior, in Secs. 
\ref{sec:above_instab}-\ref{sec:CS_strong_couple} we studied
the limit $\Gamma / \Gamma_c \rightarrow \infty$.
We found that as $\Gamma / \Gamma_c \rightarrow \infty$, Eq. (\ref{eq:LanStuCou})
reduces to Eq. (\ref{eq:LanStuNet3}), which depends 
only on coupling among oscillators individually dominated by their constant nonlinear characteristics.
Considering cluster state attractors of (\ref{eq:LanStuNet3}) we obtained the following results.

\begin{enumerate}
\item For sufficiently low values of the nonlinear frequency shift parameter ($\bar{\gamma} < \sqrt{3}$),
there is a unique, global attractor that is a single-cluster, locked state attractor.
\item For larger $\bar{\gamma}$, multiple-cluster attractors can occur, but the single-cluster,
locked state attractor continues to exist.
\item For larger $\bar{\gamma}$, two-cluster locked state
attractors exist, but locked state attractors with more than three clusters are never possible
\cite{note5}.
\item For $C=3$, regions of parameter space exist where 
two frequency quasiperiodicity can occur with periodic 
attractors arbitrarily nearby in parameter space.
\item For $C \geq 4$, $(C-1)$-frequency quasiperiodicity can occur, and periodicity and chaos
occur for parameter values near those yielding $(C-1)$-frequency quasiperiodicity.
\end{enumerate}

This work is supported by the U.S. Army Research Office grant $\#$ W911NF-12-1-0101.

% =========================================================================== appendix
\appendix
\section{Theoretical values of the critical coupling strength with
a uniformly distributed $g(\omega)$}  %\label{sec:appA}
In this appendix we summarize the theoretical results of the critical coupling strength 
$\Gamma_c$ when $g(\omega)$ is given by the uniform distribution %[Eq. (\ref{eq:Uniform})].

% -------------------------------------------------------------- gUniform_app
\begin{equation}
g(\omega) = \frac{1}{2} U(1 - |\omega|),
\end{equation}

First, we determine $\Gamma_c$ when there is no spread in $h(\alpha)$, i.e.,
$h(\alpha)=\delta(\alpha-\bar{\alpha})$, and
$\gamma=0$ for all oscillators.
When $\bar{\alpha}>0$, we have
% --------------------------------------------------------- gUnif_Gamc1
\begin{equation}
\Gamma_c^{-1} = \frac{\pi}{4} + \frac{1}{2} \tan^{-1} 
\left(  \frac{1}{2 \bar{\alpha}} \right);
\label{eq:gUnif_Gamc1}
\end{equation}

\noindent
similarly, when $\bar{\alpha}<0$, we have
% -------------------------------------------------------- gUnif_Gamc2
\begin{equation}
\Gamma_c^{-1} =  \tan^{-1} 
\left(  \frac{1}{|\bar{\alpha}|} \right).
\label{eq:gUnif_Gamc2}
\end{equation}

\noindent
For the cases when there is spread in $h(\alpha)$, we assume the same model
$h(\alpha) = (2 \delta \alpha)^{-1} U(\delta \alpha - |\alpha-\bar{\alpha}|)$. By 
denoting $\bar{\alpha}_+ = \bar{\alpha}+\delta \alpha$ and
$\bar{\alpha}_- = \bar{\alpha}-\delta \alpha$, we have
for $\bar{\alpha} > \delta \alpha$,
% ---------------------------------------------------------- gUnif_Gamc3_1
\begin{equation}
\Gamma_c^{-1} = \frac{\pi}{4} + \frac{1}{4 \delta\alpha}
\left[
\bar{\alpha}_+ \tan^{-1} \left( \frac{1}{2\bar{\alpha_+}}\right)
-\bar{\alpha}_- \tan^{-1} \left( \frac{1}{2\bar{\alpha_-}}\right)
+\frac{1}{4} \ln \left| \frac{4 \bar{\alpha}_+^2+1}
           {4 \bar{\alpha}_-^2+1} \right|
\right];
\label{eq:gUnif_Gamc3_1}
\end{equation}

\noindent
for  $\bar{\alpha} < \delta \alpha$,
% ---------------------------------------------------------- gUnif_Gamc3_2
\begin{equation}
\Gamma_c^{-1} = -\frac{1}{2 \delta\alpha}
\left[
\bar{\alpha}_+ \tan^{-1} \left( \frac{1}{\bar{\alpha_+}}\right)
-\bar{\alpha}_- \tan^{-1} \left( \frac{1}{\bar{\alpha_-}}\right)
+\frac{1}{2} \ln \left| \frac{\bar{\alpha}_+^2+1}
           {\bar{\alpha}_-^2+1} \right|
\right];
\label{eq:gUnif_Gamc3_2}
\end{equation}

\noindent
and for $|\bar{\alpha}| < -\delta \alpha$,
% ---------------------------------------------------------- gUnif_Gamc3_3
\begin{subequations}
\begin{align}
\Gamma_c^{-1} &= I_1 + I_2, \hspace{3mm} \text{where} \label{eq:gUnif_Gamc3_s1} \\
I_1 &= \frac{\pi}{4} \left( \frac{\bar{\alpha_+}}{2 \delta \alpha} \right) 
     + \frac{1}{4 \delta \alpha}
\left[
\bar{\alpha}_+ \tan^{-1} \left( \frac{1}{2\bar{\alpha_+}}\right)
+\frac{1}{4} \ln \left| 4 \bar{\alpha}_+^2+1 \right|
\right], \label{eq:gUnif_Gamc3_s2} \\
I_2 &= -\frac{1}{2 \delta\alpha}
\left[
-\bar{\alpha}_- \tan^{-1} \left( \frac{1}{\bar{\alpha_-}}\right)
-\frac{1}{2} \ln \left| \bar{\alpha}_-^2+1 \right|
\right].  \label{eq:gUnif_Gamc3_s3}
\end{align}
\label{eq:gUnif_Gamc3_3}
\end{subequations}

\noindent
Similar to the results with a Lorentzian $g(\omega)$, it can be readily shown that Eqs. 
(\ref{eq:gUnif_Gamc3_1})-(\ref{eq:gUnif_Gamc3_2}) reduce to Eqs. (\ref{eq:gUnif_Gamc1})
and (\ref{eq:gUnif_Gamc2}) in the limit $\delta \alpha \rightarrow 0$.

Next, we determine $\Gamma_c$ when there is no spread in $h(\alpha)$, i.e.,
$h(\alpha)=\delta(\alpha-\bar{\alpha})$ where $\bar{\alpha}$ is constant, 
and the nonlinear frequency parameter $\bar{\gamma}$ is a nonzero constant
for all oscillators.
For $\bar{\alpha}<0$, we know that $\bar{\gamma}$ does not affect stability 
of the state $ \langle z \rangle=0$, so $\Gamma_c$ is still
given by Eq. (\ref{eq:gUnif_Gamc2}). For
$\bar{\alpha} > 0$, we have,
by substituting $s=i \Omega$ into the final expression after integration
in Eq. (\ref{eq:Gaminv}),
and introducing 
$\eta_\pm = \Omega \pm 1+\bar{\alpha}\bar{\gamma}$, that
$\Gamma_c$ and $\Omega$ are to be given by the solution of the following pair of equations,
% ---------------------------------------------------------- gUnif_NL_Gamc1
\begin{subequations}
\begin{align}
-4  \Gamma_c^{-1} &= \frac{\bar{\gamma}}{2} \ln \left|
\frac{\eta_+^2 (\eta_-^2+4\bar{\alpha}^2)}{\eta_-^2 (\eta_+^2+4\bar{\alpha}^2)} \right|
-\left[ \pi + \tan^{-1}\left( \frac{\eta_+}{2\bar{\alpha}} \right) 
-\tan^{-1}\left( \frac{\eta_-}{2\bar{\alpha}} \right)  \right], \label{gUnif_NL_Gamc1_s1} \\
0 &= \bar{\gamma} \left[ \pi - \tan^{-1}\left( \frac{\eta_+}{2\bar{\alpha}} \right) 
+\tan^{-1}\left( \frac{\eta_-}{2\bar{\alpha}} \right)  \right] + 
\frac{1}{2} \ln \left|
\frac{\eta_+^2 (\eta_+^2+4\bar{\alpha}^2)}{\eta_-^2 (\eta_-^2+4\bar{\alpha}^2)} \right|.
\label{eq:gUnif_NL_Gamc1_s2}
\end{align}
\label{eq:gUnif_NL_Gamc1}
\end{subequations}

\noindent
It can be easily checked from (\ref{eq:gUnif_NL_Gamc1}) that
$\Gamma_c$ reduces to
(\ref{eq:gUnif_Gamc1}) when $\bar{\gamma}  \rightarrow 0$
(Note $\Omega \rightarrow 0$ in this limit).

% =========================================================================

% ========================================================================


\begin{thebibliography}{99}
\bibitem{note1}
In general, however, we note that $\omega$ and $\alpha$ can be expected to 
have correlations; e.g., even if the physical parameter vector ${\bf p}$
has dimension $n \geq 2$ and the variation of components of ${\bf p}$  are uncorrelated, 
we can still expect that the particular functional dependences, Eq. (\ref{eq:para}),
of the oscillator parameters on the physical parameters will induce correlations
between $\omega$ and $\alpha$.
\bibitem{note2}
In addition to the situation of many coupled Landau-Stuart equations, oscillator
death occurs very generally for many types of coupled oscillator 
situations, including coupling between only a few oscillators (e.g., two).
As an example of one of the many references on this topic, we note the
recent paper, Ref. \cite{Suarez}, and references therein.
\bibitem{note3}
Equation (\ref{eq:Gam_c1}) [together with a transformation of the form of the interaction
term to that of Eq. (\ref{eq:LanStu_oldw})] can also be used to
generalize previous work of Ref. \cite{Daido} on the ``aging transition''
to include dispersion of the natural frequencies.
\bibitem{note4}
It can be shown that when $h(\alpha) = \delta (\alpha-1)$,
$\Gamma_c^{-1} = \pi g(0)$ for any general unimodal
frequency distribution $g(\omega)$ symmetric about $\omega=0$
(see Ref. \cite{Matthews2}).
\bibitem{note5}
We do not yet know whether $C=3$ locked state attractors exist, but we have so far
not observed them in our simulations.
\end{thebibliography}
\end{document}